\documentclass[format=acmsmall, review=false, screen=true]{acmart}

\usepackage{algorithm}
\usepackage{algorithmicx}
\usepackage[disable]{todonotes} 
\usepackage[noend]{algpseudocode}

\algnewcommand{\IfInline}[1]{\State\algorithmicif\ \, #1\ \, \algorithmicthen \, }
\algnewcommand{\EndIfInline}{\unskip\ }

\usepackage{amsthm}
\usepackage{amssymb}
\usepackage{booktabs} 
\usepackage{framed}
\usepackage{url}
\usepackage{color}
\usepackage{xcolor}
\usepackage{subfig}
\usepackage[skip=1pt]{caption}
\usepackage{tabularx}
\usepackage{rotating}
\usepackage{url}
\usepackage{hyperref}
\usepackage{amsmath}
\usepackage{multirow}
\usepackage{enumerate}

\algdef{SE}[DOWHILE]{Do}{doWhile}{\algorithmicdo}[1]{\algorithmicwhile\ #1}

\newtheorem{remark}{Remark}

\newtheorem{definition}{Definition}
\newcommand{\cC}{{\mathcal C}}
\newcommand{\cE}{{\mathcal E}}

\newcommand{\cG}{{\mathcal G}}
\newcommand{\cH}{{\mathcal H}}
\newcommand{\cK}{{\mathcal K}}
\newcommand{\cL}{{\mathcal L}}

\newcommand{\cV}{{\mathcal V}}

\newcommand{\tvert}{q}

\newcounter{alg}
\begin{document}


\title{Scalable Pattern Matching in Metadata Graphs \\ via \textit{Constraint Checking}}

\author{Tahsin Reza}
\author{Hassan Halawa}
\author{Matei Ripeanu}
\affiliation{%
  \institution{\\University of British Columbia}
  \city{Vancouver}
  \state{BC}
  }
\email{treza@ece.ubc.ca}
\email{hhalawa@ece.ubc.ca}
\email{matei@ece.ubc.ca}
\author{Geoffrey Sanders}
\author{Roger A. Pearce}
\affiliation{%
  \institution{\\Lawrence Livermore National Laboratory}
  \city{Livermore}
  \state{CA}
  }
\email{sanders29@llnl.gov}
\email{rpearce@llnl.gov}

\begin{abstract}
Pattern matching is a fundamental tool for answering complex graph queries. Unfortunately, existing solutions have limited capabilities: they do not scale to process large graphs and/or support only a restricted set of search templates or usage scenarios. Moreover, the algorithms at the core of the existing techniques are not suitable for today's graph processing infrastructures relying on horizontal scalability and shared-nothing clusters, as most of these algorithms are inherently sequential and difficult to parallelize.

We present an algorithmic pipeline that bases pattern matching on \emph{constraint checking}. The key intuition is that each vertex and edge participating in a match has to meet a set of constraints implicitly specified by the search template. These constraints can be verified independently, and typically, are less expensive to compute than searching the full template. The pipeline we propose generates these constraints and iterates over them to eliminate all the vertices and edges that \emph{do not} participate in any match, thus reducing the background graph to a subgraph which is the union of all template matches - the \textit{complete set} of all vertices and edges that participate in at least one match. Additional analysis can be performed on this annotated, reduced graph, such as full match enumeration, match counting, or computing vertex/edge centrality. Furthermore, a \emph{vertex-centric} formulation for constraint checking algorithms exists, and this makes it possible to harness existing high-performance, vertex-centric graph processing frameworks.

This technique, \textit{(i)} enables highly scalable pattern matching in metadata (labeled) graphs; \textit{(ii)} supports arbitrary patterns with 100\% precision; \textit{(iii)} enables trade-offs between precision and time-to-solution, while always selects all vertices and edges that participate in matches, thus offering 100\% recall; and \textit{(iv)} supports a set of popular data analytics scenarios. We implement our approach on top of HavoqGT, an open-source asynchronous graph processing framework, and demonstrate its advantages through strong and weak scaling experiments on massive scale real-world (up to 257 billion edges) and synthetic (up to 4.4 trillion edges) labeled graphs, respectively, and at scales (1,024 nodes / 36,864 cores), orders of magnitude larger than used in the past for similar problems.

This paper serves two purposes: First, it synthesises the knowledge accumulated during a long-term project ~\cite{Reza:2017:IEEECluster:PM.8048872,Reza:2018:SC:PM,Nicolas:2018:IA3:PM:8638389}. Second, it presents new system features, usage scenarios, optimizations, and comparisons with related work, that strengthen the confidence that pattern matching based on iterative pruning via constraint checking is an effective and scalable approach in practice.  The new contributions include: \textit{(i)} We demonstrate the ability of the constraint checking approach to 
efficiently support two additional search scenarios that often emerge in practice, \emph{interactive incremental search} and \emph{exploratory search}. 
\textit{(ii)} We empirically compare our solution with two additional state-of-the-art systems, Arabsque~\cite{Teixeira:2015:ASD:2815400.2815410} and TriAD~\cite{Gurajada:2014:TDS:2588555.2610511}.
\textit{(iii)} We show the ability of our solution to accommodate a more diverse range of datasets with varying properties, e.g., scale, skewness, label distribution and match frequency.
\textit{(iv)} We introduce or extend a number of system features (e.g., work aggregation, load balancing, the ability to cap the generated traffic) and 
design optimizations, and demonstrate their advantages with respect to improving performance and scalability.
\textit{(v)} We present bottleneck analysis and insights into artifacts that influence performance.
\textit{(vi)} We present a theoretical complexity argument that motivates the performance gains we observe.
\end{abstract}

\ccsdesc[500]{Computer systems organization~Distributed architectures}
\ccsdesc[500]{Information systems~Data mining}
\ccsdesc[500]{Mathematics of computing~Graph algorithms}

\keywords{Pattern matching, Subgraph isomorphism, Graph processing, Distributed computing}

\maketitle

\renewcommand{\shortauthors}{T. Reza et al.} 

\section{Introduction}
\label{sxn:introduction}

\noindent
Pattern matching in labeled graphs, that is, finding subgraphs that \emph{match} a small \textit{template graph} within a large \textit{background graph}, is fundamental to graph analysis and has applications in social network analysis~\cite{Fan:2013:DTG:2536258.2536263,Fan:2010:GPM:1920841.1920878, Tong:2007:FBP:1281192.1281271,Gupta:2014:RTR:2733004.2733010},  bioinformatics~\cite{Alon:2008:BNM:1388083.1388101}, anomaly and fraud detection~\cite{Iyer:2018:ASAP:222637}, program analysis~\cite{Lo:2009:CSB:1557019.1557083}, as well as in various machine learning contexts~\cite{Henderson:2012:RSR:2339530.2339723, Grover:2016:NSF:2939672.2939754}. A \emph{match} can be broadly categorized as either \textit{exact} - i.e., there is a bijective mapping between the vertices/edges in the template and those in the matching subgraph, or \textit{approximate} - the template and the match are just similar by some defined similarity metric ~\cite{Bunke:1983:IGM:2305869.2306079, CONTE.2004.TYGMPR.doi:10.1142.S0218001404003228, Zhang:2010:SSI:1920841.1920988}. 

\begin{figure}[!h]
\centering
\includegraphics[width=3.0in]{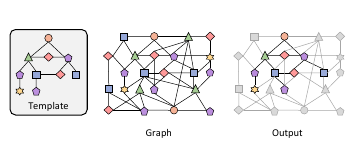} 
\captionsetup{font=footnotesize}
\caption{An example of a background graph $\cG$ (center), template $\cG_0$ (left) and the output - the solution subgraph $\cG^*$ after vertex and edge elimination (right). The output is a refined set of vertices and edges that participate in at least one subgraph $\cH$ that matches $\cG_0$. Here (and in the rest of the paper), vertex metadata are presented as colored shapes. The eliminated vertices and edges are colored solid grey. (Reused from~\cite{Reza:2018:SC:PM}.)}
\label{fig:example_0011}
\end{figure}

Unfortunately, existing pattern matching solutions (related work in \S\ref{snx:related_work}) have limited capabilities: \textit{(i)} they do not scale to the massive graphs with hundreds of billions of edges commonly mined nowadays; \textit{(ii)} they often support only a restricted set of search templates or usage scenarios; and \textit{(iii)} they rely on algorithms that are not suitable for implementation on top of today's graph processing infrastructures which aim at horizontal scalability and shared-nothing clusters, as most of these algorithms are inherently sequential and difficult to parallelize \cite{Ullmann:1976:ASI:321921.321925, P.Cordella:2004:GIA:1018035.1018377, Mckay:2014:PGI:2562352.2562750}.  

We propose a new algorithmic pipeline based on \emph{constraint checking}. This approach is motivated by viewing the search template as specifying a set of constraints the vertices and edges that participate in a match must meet. The pipeline iterates over these constraints to eliminate \textit{all} and \textit{only} the vertices and edges that \emph{do not} participate in any match. The intuition for the effectiveness of this technique stems from four key observations:
\\

\textit{(i)} First, the traditionally used \emph{graph exploration} techniques~\cite{Ullmann:1976:ASI:321921.321925, P.Cordella:2004:GIA:1018035.1018377, Shang:2008:TVH:1453856.1453899, Han:2014:CGE:2592798.2592799} generally attempt to \emph{enumerate} all matches through explicit search. When an exploration branch fails, it has to be marked invalid and ignored in the subsequent steps. In the same vein as past works that use graph pruning~\cite{Lulli.2017.TPDS.7515231,Zhou2012} or, more generally, input reduction~\cite{Kusum:2016:EPL:2907294.2907312}, we observe that it is much cheaper to focus on eliminating the vertices and edges that do not meet the label and topological constraints introduced by the search template.
\textit{One key contribution of this work
is a pruning-based solution that eliminates  
\textit{all} and \textit{only} the vertices and edges that do not participate in any match, limits the exponential growth of the algorithm state, scales to massive graphs and distributed memory machines with a large number of processing elements, and supports arbitrary search templates}.
The result of pruning is the \textit{complete set} of all vertices and edges that participate in at least one match, with \emph{no false positives} or \emph{false negatives}. Fig.~\ref{fig:example_0011} illustrates the general idea using an example graph and a search template.
\\

\textit{(ii)} Second, we observe that, \emph{full match enumeration} is not the most efficient avenue to support many high-level graph analysis scenarios. Depending on the final goal of the user, pattern matching problems fall into a number of categories which include: \textit{(a)} determining if a match exists in the background graph (yes/no answer), \textit{(b)} selecting all the vertices and edges that participate in matches, \textit{(c)} ranking these vertices or edges based on their \emph{centrality with respect to the search template}, e.g., the frequency of their participation in matches, \textit{(d)} counting/estimating the total number of matches, or \textit{(e)} enumerating all distinct matches in the background graph. The traditional approach~\cite{Ullmann:1976:ASI:321921.321925,P.Cordella:2004:GIA:1018035.1018377} is to first enumerate the matches (category \textit{(e)} above) and to use the result to answer \textit{(a) -- (d)}. However, this approach is limited to small background graphs or is dependent on a low number of near and exact matches within the background graph (due to exponential growth of the algorithm state). Our experiments suggest that that a pruning-based approach is not only a practical solution to scenarios \textit{(a) -- (d)} (and to other pattern matching related analytics) but also an efficient path towards full match enumeration in large graphs.
There are three reasons for the effectiveness of this approach: First, the pruned graph can be multiple orders of magnitude smaller than the background graph, and existing high-complexity enumeration routines thus become applicable. Second, our pruning techniques collect additional key information to accelerate match enumeration - for each vertex in the pruned graph, our algorithms build a list of its possible match(es) in the template. Lastly, the intermediate algorithm state is much smaller. 
\\

\textit{(iii)} Third, such pruning approach lends itself well to developing a \emph{vertex-centric} algorithmic solution, and this makes it possible to harness existing high-performance, vertex-centric frameworks (e.g., GraphLab~\cite{Gonzalez:2012:PDG:2387880.2387883}, Giraph~\cite{Giraph.001} or HavoqGT~\cite{Pearce:2014:FPT:2683593.2683654}). In our vertex-centric formulation for pruning, a vertex must satisfy two types of constraints, namely, \emph{local} and \emph{non-local}, to possibly be part of a match. \emph{Local constraints} involve only the vertex and its neighborhood: a vertex in an exact match needs to \textit{(a)} match the label of a corresponding vertex in the template, and \textit{(b)} have edges to vertices labeled as prescribed in the adjacency structure of this corresponding vertex in the search template.
\emph{Non-local constraints} are topological requirements beyond the immediate neighborhood of a vertex (e.g., that the vertex must be part of a cycle). We describe how these constraints are generated, and our algorithmic solution to verify them in \S\ref{sxn:algs}. 
\\

\textit{(iv)} Finally, decomposing the search template in a set of constraints enables, in addition to \textit{exact matching}, efficiently supporting a number of additional usage scenarios. These include: \textit{(a)} trade-offs between precision and time-to-solution as search can be stopped early after checking only a subset of the constraints - leading to lower precision in the solution set (\cite{Reza:2018:SC:PM}, \S5E); \textit{(b)} \textit{incremental searches} - an interactive search scenario where the user incrementally updates the search template (while the system takes advantage of the existence of common constraint(s) in the past and the updated search template, to offer fast response time) (\S\ref{sxn:evl_usage_scenarios_8}); 
\textit{(c)} \textit{exploratory search} - a search scenario where the user presents an over-constrained search template that may not have any match, and the system finds the `nearest' matches (e.g., the ones that satisfy most of the constraints of the original search template) (\S\ref{sxn:evl_usage_scenarios_8}); and finally, \textit{(d)} \textit{approximate searches} based on \emph{edit-distance}~\cite{Bunke:1983:IGM:2305869.2306079,Zhang:2010:SSI:1920841.1920988,Reza.2020.SIGMOD.10.1145/3318464.3380566}.
\\

\textbf{\textit{Contributions.}} 
This paper serves two goals: first, it is a synthesis of an ongoing long-term project~\cite{Reza:2017:IEEECluster:PM.8048872,Reza:2018:SC:PM,Nicolas:2018:IA3:PM:8638389, Reza.2020.SIGMOD.10.1145/3318464.3380566,Reza.2020.GRADES.10.1145/3398682.3399166}; and second, it presents new system features, usage scenarios, empirical experiments, and comparisons with related projects, that strengthen the confidence that pattern matching based on iterative pruning via constraint checking is an effective and scalable approach. 
\\

\textit{Summary of Previously Published Work.} 
We have introduced the \emph{constraint checking} approach and discussed the opportunities it presents to pattern matching in large-scale graphs in a preliminary study~\cite{Reza:2017:IEEECluster:PM.8048872}.  This study focused on the exact matching scenario only and a restricted set of search templates: \emph{acyclic} or
\emph{edge-monocyclic} with unique vertex labels (see \S\ref{snx:prelm} for definitions). A key contribution of this preliminary study proving that, for some templates, only in expensive local constraint checking is sufficient for a precise solution. Following this initial investigation, we introduced PruneJuice~\cite{Reza:2018:SC:PM}, a distributed system for exact pattern matching that is: \textit{generic} - no restrictions on the set of patterns supported, \textit{precise} - no false positives and \textit{offers 100\% recall} - retrieves all matches, \textit{efficient} - small algorithm state ensuring low generated network traffic, and \textit{scalable} - able to process graphs with up to trillions of edges on tens of thousands of cores. Here, also the focus was the exact matching scenario only. Strong and weak scaling experiments using massive background graphs and scaling to up to 1,024 nodes (36,864 cores), confirmed the scalability of this approach. We demonstrated that, depending on the input, pruning leads to a \emph{solution subgraph} that can be orders of magnitude smaller than the original background graph, which enables match enumeration and counting in massive graphs.  While this study used simple heuristics to select and order constraints, we have demonstrated the effectiveness of advanced heuristics for constraint selection and ordering in the context of a shared memory implementation in~\cite{Nicolas:2018:IA3:PM:8638389}. At the level of system architecture and implementation, the following key design ingredients make our system successful: asynchronicity, aggressive vertex and edge elimination while harnessing massive parallelism, intelligent work aggregation to ensure low message overhead, and lightweight per-vertex state.  While in this paper  we focus on exact matching and closely associated scenarios, we have shown the finer granularity a constraint-based approach facilitates, can enable a version of approximate matching based on edit-distance~\cite{Reza.2020.SIGMOD.10.1145/3318464.3380566}.
\\

\textit{Summary of New Contributions.} 
While one goal of this manuscript is to synthesize and organize the experience we have acquired during this project, it also includes an entirely new evaluation section, and original material as follows. 

\begin{itemize}

\item[\textit{(i)}]
\textit{Demonstrating Support for a Diverse Set of Graph Analytics Scenarios} (\S\ref{sxn:evl_usage_scenarios_8}).  We show the ability of our approach to efficiently support an \emph{exploratory search} scenario where the user starts form an over-constrained search template the system progressively relaxes the search until matches are found.  We also present an overview of an \emph{interactive incremental search} which supports the following usage scenario: the user may not know exactly what (s)he is looking for, and based on returned results, will incrementally revise the search template, possibly multiple times. Two high-level features presented by the exact matching pipeline are essential to support these additional scenarios: First, decomposing the search template into a set of constrains enables partial result reuse; second, a prunning based approach makes it natural to focus on the part of the data of interest thus improving locality and reducing generated network traffic. We discuss the unique design optimizations enabled by the constraint checking approach to support these scenarios, and demonstrate their effectiveness through experiments on massive graphs.
\\

\item[\textit{(ii)}] 
\textit{Comparison with State-of-the-Art Systems} (\S\ref{sxn:compare_others}).
We extensively compare our work with two additional state-of-the art distributed solutions, Arabesque~\cite{Teixeira:2015:ASD:2815400.2815410} (\S\ref{sxn:compare_arabesque}) and TriAD~\cite{Gurajada:2014:TDS:2588555.2610511} (\S\ref{sxn:compare_triad}), in multiple scenarios (match enumeration and counting), and using labeled and unlabeled templates, and multiple  real-world graphs. These experiments demonstrate the significant advantages our system offers when handling large graphs, and complex labeled or unlabeled patterns. 
\\

\item[\textit{(iii)}] 
\textit{Additional System Optimizations and Experiments using New Datasets.} 
We have further added a number of optimizations aimed at enhancing performance, scalability, robustness and efficiency. These include: \emph{work aggregation} - a light-weight yet highly effective technique to prevent relaying duplicate messages (\S\ref{snx:di}); load balancing (\S\ref{sxn:load_balancing_evaluation}); 
and the ability to control the processing rate to lower memory pressure. 
We demonstrate that these techniques offer multitude of performance gains as well as robustness when processing at a massive scale (\S\ref{snx:optimization_evaluation}, \S\ref{sxn:load_balancing_evaluation} and \cite{Reza:2018:SC:PM}).  The extended evaluation includes new real-world graphs and three synthetic graphs with varying topology, skewness and degree distribution; and several new search templates, both labeled and unlabeled (\S\ref{sxn:evl_strong_scaling}).
\\

\item[\textit{(iv)}] 
\textit{Bottleneck Analysis and Insights into Artifacts that Influence Performance.} 
We present experiments that uncover artifacts that influence performance along multiple axes. We explore the artifacts that cause load imbalance (\S\ref{sxn:load_balancing_evaluation}); 
and we investigate the influence of search template and background graph properties (e.g., label distribution and topology) on runtime performance (\S\ref{sxn:evl_template_sensitivity_analysis} and \S\ref{sxn:evl_graph_sensitivity_analysis}).
\\

\item[\textit{(v)}] 
\textit{Complexity Analysis} (\S\ref{sxn:complexity_analysis}). We present runtime, message count, and storage complexity of the core constraint checking algorithms. We also present a theoretical complexity argument that motivates the performance gains we observe.

\end{itemize}


%
%
\section{Related Work}
\label{snx:related_work}

\noindent
The volume of related work on graph processing in general~\cite{Malewicz:2010:PSL:1807167.1807184,Giraph.001,Gonzalez:2012:PDG:2387880.2387883,Gonzalez:2014:GGP:2685048.2685096,Sundaram:2015:GHP:2809974.2809983,Hong:2015:PFD:2807591.2807620}, and on pattern matching algorithms in particular~\cite{Ullmann:1976:ASI:321921.321925,P.Cordella:2004:GIA:1018035.1018377,Mckay:2014:PGI:2562352.2562750,Zhu.2011.d6825c67775f402ebc36f498d218a971,Berry.2007.4228413,Fan:2013:DTG:2536258.2536263}, is humbling. We summarize closely related work in Table \ref{table:related_work_distributed_pattern_matching}.

\subsection{General Algorithmic Approaches for Exact Pattern Matching}
Early work on graph pattern matching mainly focused on solving the graph isomorphism problem \cite{Ullmann:1976:ASI:321921.321925}. The well-known Ullmann's algorithm \cite{Ullmann:1976:ASI:321921.321925} and its improvements (in terms of join order, pruning strategies and space complexity), e.g., VF2~\cite{P.Cordella:2004:GIA:1018035.1018377} and QuickSI~\cite{Shang:2008:TVH:1453856.1453899}, belong to the family of \textit{tree-search} based algorithms. Ullman proposed a backtracking algorithm which finds exact matches by incrementing partial solutions and uses heuristics to prune unprofitable paths. VF2 improves the time and space complexity over Ullman's algorithm. 
The algorithm uses a heuristic that is based on the analysis of the vertices adjacent to vertices that have been included in a partial solution. The VF2 algorithm is known to be robust and performs well in practice, 
and consecutively has been included in the popular Boost Graph Library (BGL)~\cite{Boost.Graph.Library.001}. A recent effort, Turbo\textsubscript{ISO}~\cite{Han:2013:TIT:2463676.2465300} is considered to be the 
most optimized among the tree-search based sequential subgraph isomorphism techniques. (Note that the pattern search can be performed in a depth-first or a breadth-first manner. The na\"ive pattern matching technique recursively searches the full template from each vertex in the background graph in a depth-first manner. The tree-search algorithms are merely optimizations of this depth-first search technique.) 

For large graphs, a tree search may fail midway and needs to backtrack, hence, this technique can be expensive. Efficient distributed implementation of this approach is difficult for a number of reasons: existing algorithms are inherently sequential and difficult to parallelize. 
Furthermore, a key limitation of this technique is that the number of possible join operations (the process of adding a graph edge to an intermediate match) is combinatorially large; which makes its application to generic patterns and massive graphs, with billions or trillions of edges, impractical. 
Also, the above algorithms use heuristics for join order selection~\cite{Han:2013:TIT:2463676.2465300}, as a result, often the performance is sensitive to the graph topology, label frequency, and relies on expensive preprocessing for join order optimization, such as sorting the neighbor vertices by degree.  

Perhaps the best known exact matching algorithm that does not belong to the family of tree-search algorithms is Nauty due to McKay \cite{Mckay:2014:PGI:2562352.2562750}, which is based on \textit{canonical labeling} of the graph. This approach, however, has high preprocessing overhead. Nauty can perform verification for isomorphism in $O(n^2)$ time (where $n$ is the number of vertices in the background graph), however, transforming arbitrary input graphs to the canonical form requires exponential time~\cite{Miyazaki.1997.003}.

In the same spirit as database indexing, \textit{subgraph indexing} (i.e., indexing of frequent subgraph structures) is an approach attempted in order to reduce the number of join operations (between subgraph structures) 
and to lower query response time, e.g., SpiderMine~\cite{Zhu.2011.d6825c67775f402ebc36f498d218a971}, R-Join~\cite{Cheng.2008.4497500}, C-Tree~\cite{Singh.2006.ICDE.CTree.1617406}, SAPPER~\cite{Zhang:2010:SSI:1920841.1920988}, TriAD~\cite{Gurajada:2014:TDS:2588555.2610511}, and the contributions by Sun et al.~\cite{Sun:2012:ESM:2311906.2311907} and Gao et al.~\cite{Gao.2014.6816681}. Unfortunately, for a billion edge graph, this approach is infeasible to generalize: First, searching frequent subgraphs in a large graph is notoriously expensive. Second, depending on the topology of the search template(s) and the background graph, the size of the index is often superlinear relative to the size of the graph~\cite{Sun:2012:ESM:2311906.2311907}.  

\subsection{Distributed Pattern Matching Solutions}
We review a number of projects that offer pattern matching on a shared-nothing architecture aiming either to reduce time-to-solution or to scale to search in large background graphs. 
Table~\ref{table:related_work_distributed_pattern_matching} summarizes the key differentiating aspects and the scale achieved. Below we group the contributions into exact and approximate matching categories. 
\\

\begin{table*}[!t]
\footnotesize
\setlength{\tabcolsep}{0.1pt}
\renewcommand{\arraystretch}{1.2}
\captionsetup{font=footnotesize}
\caption{Comparison of past work on distributed pattern matching.  The table highlights the characteristics of the solution presented (exact vs. approximate matching), its implementation infrastructure, and summarizes the details of the largest scale experiment performed. We highlight the fact that \emph{our solution is unique in terms of demonstrated scale, ability to perform exact matching, and ability to retrieve all matches}.}
\label{table:related_work_distributed_pattern_matching}
\centering
\resizebox{\textwidth}{!}{
\begin{tabular}{lcccccccccccc}
\hline
\multirow{2}{*}{Contribution} & 
\multirow{2}{*}{Model} & 
Framework/ & Match 
& Max. Query 
& Metadata 
& \#Compute 
& Max. Real 
& Max. Synthetic 
\\
&&Language&Type&Size&&Nodes&Graph&Graph\\
\hline
Arabesque \cite{Teixeira:2015:ASD:2815400.2815410} & Tree-search & Spark & Exact 
& 10 edges & N/A & 20 & 887M edges & N/A \\
QFrag \cite{Serafini:2017:QDG:3127479.3131625} & Tree-search & Spark & Exact 
& 7 edges & Real & 10 & 117M edges & N/A \\
Fractal \cite{Dias.2019.SIGMOD.10.1145/3299869.3319875} & Tree-search & Spark & Exact 
& 10 edges & Real & 10 & 44M edges & N/A \\
PGX.D/Async \cite{Roth:2017:PSD:3078447.3078454} & Async. DFS & Java/C++ & Exact 
& 4 edges & Synthetic & 32 & N/A & 2B edges (Unif. rand.) \\
G-Miner \cite{Chen:2018:GET:3190508.3190545} & Tree-search & C++ & Exact 
& 4 edges & N/A & 15 & 1.8B edges & N/A \\
Sun et al.
\cite{Sun:2012:ESM:2311906.2311907} & Subgraph Indexing & C\#{}.Net4 & Exact 
& 30 edges & Synthetic & 12 & 16.5M edges & 4B vertices \\
Plantenga \cite{Plantenga:2013:ISI:2416443.2416465} & Tree-search & Hadoop & Approximate 
& 4-Clique & Real & 64 & 107B edges & R-MAT Scale 20 \\
SAHAD \cite{Zhao.2012.SAHAD.6267876} &
Color-coding & Hadoop & Approximate & 12 vertices & Synthetic & 40 & N/A & 269M edges 
\\
FASCIA \cite{Slota:2014:IPDPS:SM14b} &
Color-coding & MPI & Approximate & 12 vertices & N/A & 15 & 117M edges & 1M edges (Erd\H{o}s-R\'enyi) 
\\
Chak. et al.~\cite{Chakaravarthy.2016.IPDPS.7515996} & 
Color-coding & MPI & Approximate & 10 vertices & N/A & 512 (BG/Q) & 2.7M edges & R-MAT
\\
Gao et al.
\cite{Gao.2014.6816681} & Subgraph Indexing & Giraph & Approximate 
& 50 vertices  & Synthetic & 28 & 3.7B edges & N/A \\
Ma et al.
\cite{Ma:2012:DGP:2187836.2187963} & Graph Simulation & Python & Approximate 
& 15 vertices & Type only & 16 & 5.1M edges & 100M vertices \\
Fard et al.
\cite{Fard.2013.6691601} & Graph Simulation & GPS & Approximate & N/A & N/A 
& 8 & 300M edges & N/A \\
ASAP \cite{Iyer:2018:ASAP:222637} & Neighborhood Sampling & Spark & Probabilistic 
& 6 edges & N/A & 16 & 3.7B edges & N/A \\
Yuan et al. \cite{Yuan:2015:EDS:2777067.2777084} & Tree-search/Join & Java & Exact 
& 17 edges & N/A & 17 & 1.4B edges & 64M vertices \\
\hline
\end{tabular}
}
\end{table*}

\textit{\textbf{Solutions Offering Exact Matching.}}
Arabesque~\cite{Teixeira:2015:ASD:2815400.2815410} is a distributed framework offering precision and recall guarantees, implemented on top of Apache Spark~\cite{Spark.001} and Giraph~\cite{Giraph.001}. Arabesque provides an API based on the Think Like an Embedding (TLE) paradigm, to express graph mining algorithms 
and a Bulk Synchronous Parallel (BSP) implementation of the embedding (pattern) search engine (which follows the tree-search approach for match enumeration and counting). Arabesque replicates the input graph on all worker nodes, hence, the largest graph scale it can support is limited by the size of the memory of a single node (the implementation also exploits HDFS storage to maintain partially computed embeddings). Through evaluation using several real-world graphs, Teixeira et al.~\cite{Teixeira:2015:ASD:2815400.2815410} showed Arabesque's superiority over other two key systems: G-Tries~\cite{Ribeiro:2014:GDS:2589412.2589422} and GRAMI~\cite{Elseidy:2014:GFS:2732286.2732289}. In \S\ref{sxn:compare_arabesque}, 
we directly compare our work with Arabesque.

QFrag~\cite{Serafini:2017:QDG:3127479.3131625} is a general purpose exact pattern matching system, built on top of Arabesque. Similar to Arabesque, QFrag assumes that the entire graph fits in the memory of each compute node and uses data replication to enable search parallelism. QFrag employs a sophisticated load balancing strategy to reduce time-to-solution.  In QFrag, each replica runs an instance of the tree-search based pattern enumeration algorithm, Turbo\textsubscript{ISO}~\cite{Han:2013:TIT:2463676.2465300} (an improvement of Ullmann's algorithm~\cite{Ullmann:1976:ASI:321921.321925}). Through evaluation, the authors demonstrated QFrag's performance advantages over two other distributed pattern matching systems: TriAD~\cite{Gurajada:2014:TDS:2588555.2610511}, an MPI-based distributed RDF~\cite{RDF.W3C.003} engine based on an asynchronous distributed join algorithm, and GraphFrames~\cite{Dave:2016:GIA:2960414.2960416, GraphFrames.001}, a graph processing library for Apache Spark, also based on distributed join operations. Although Arabesque and QFrag outperform most of their competitors in terms of time-to-solution, they replicate the entire graph in the memory of each compute node, which limits their applicability to relatively small graphs. In \S\ref{sxn:compare_qfrag} and \S\ref{sxn:compare_triad}, 
we present direct comparison of our work with QFrag and TriAD, respectively. 

Similar to Arabesque, G-Miner~\cite{Chen:2018:GET:3190508.3190545} offers a high-level API for implementing graph mining algorithms; however, its applicability seems to be restricted to a limited scenarios as evaluation results were presented only for counting triangles and small cliques. A new framework Fractal~\cite{Dias.2019.SIGMOD.10.1145/3299869.3319875}, also based on the TLE abstraction, addresses several limitations of Arabesque to offer improved performance and memory efficiency.

PGX.D/Async~\cite{Roth:2017:PSD:3078447.3078454} is a distributed system by Oracle Labs offering exact matching. It relies on asynchronous depth-first traversal for match enumeration. PGX.D/Async offers an MPI-based implementation and incorporates a flow control mechanism with a deterministic guarantee of search completion under a finite amount of memory; however, compared to our work, was demonstrated at a much smaller scale, in terms of graph sizes and number of compute nodes.

Sun et al.~\cite{Sun:2012:ESM:2311906.2311907} present an exact subgraph matching solution which follows the tree-search and join approach and demonstrate it on large synthetic graphs, using larger search templates than in~\cite{Plantenga:2013:ISI:2416443.2416465}, yet not on real-world graphs. 
Also, the authors mentioned that they terminate the search after the match-count have reached a predefined threshold which was set to 1,024 in their experiments (i.e., does not offer recall guarantees).

The Graph database Engine for Multithreaded Systems (GEMS), a framework for implementing RDF databases on distributed platforms~\cite{Castellana.2015.IEEEComputer.7063171} supports SPARQL and GraQL queries, as well as user queries written in C++. GEMS exploits the Partitioned Global Address Space (PGAS) programming model: PGAS exposes the distributed memory of cluster nodes with a shared memory abstraction~\cite{Morari.2015.Cluseter.7307591}. The GEMS runtime engine manages parallelism supporting thousands of lightweight parallel tasks per node. The authors have shown GEMS scalability using an RDF dataset with up to 10 billion triples on up to 128 compute nodes on an HPC platform~\cite{Morari.2015.book.bigdata.aaa}. GraQL is a query language for GEMS that address a number of limitations of typical relational and native graph abstractions for supporting the Property Graph model~\cite{Chavarría.2016.IPDPSW.7530036}. 
\cite{Choudhury.2015.EDBT.SACP} also explored the problem of pattern matching in streaming graphs within GEMS and evaluated their solution using acyclic search templates.

A number of projects on high-performance pattern matching, primarily target genome sequencing/assembly problems. In~\cite{LOW20071007}, the authors present a solution for multiple sequence alignment for handling a large number of protein sequences on a distributed platform with the mesh topology. The technique divides the computational load among compute nodes as opposed to dividing a protein sequence among compute nodes, with the goal of maximizing throughput. In~\cite{Liu.2013.TPDS.6226376}, a parallel sequence assembly solution for multi-core shared memory platforms is presented: the solution incorporates a suffix array based data structure and can efficiently handle a large number of \emph{reads}. In~\cite{Yin.2016.JCO.s10878-015-9940-4}, an improved technique for median computation in gene matching (or difference computation) is proposed: the solution models the problem in a manner which enables harnessing subgraph matching and search space reduction, towards offering efficiency. \cite{Makkar.2017.HiPC.8287730} presents a GPU-based solution for triangle counting in dynamic graphs. The authors presented an inclusion-exclusion formulation for the problem that correctly counts the number of triangles; the solution also improves complexity bounds over prior approaches. \cite{Green.2018.HPEC.8547581} presents Logarithmic Radix Binning for triangle counting; the work presents a multi-threaded and vectorized solution which also eliminates branch misprediction.
\\

\textit{\textbf{Solutions Targeting Approximate Matching.}}
The best demonstrated scale is offered by~\cite{Plantenga:2013:ISI:2416443.2416465}: a MapReduce implementation of the \emph{walk}-based algorithm for identifying \emph{type-isomorphic} (approximate) matches, originally proposed in \cite{Berry.2007.4228413}. 
Plantenga introduced the idea of \emph{walk-level constraints} to type-isomorphism - the added constraints are expected to reduce the search space of candidate walks

SAHAD \cite{Zhao.2012.SAHAD.6267876} is a MapReduce implementation of the \emph{color-coding} algorithm~\cite{Alon:2008:BNM:1388083.1388101} originally developed for approximating the count of tree-like patterns (a.k.a. \emph{treelet}) in protein-protein interaction networks. SAHAD follows a hierarchical sub-template explore-join approach. Its application was presented only on small graphs with up to $\sim$300M edges. FASCIA~\cite{Slota:2014:IPDPS:SM14b} is also a color-coding based solution for approximate treelet counting, whose MPI-based implementation offers superior performance to SAHAD. Chakaravarthy et al.~\cite{Chakaravarthy.2016.IPDPS.7515996} extended the color-coding algorithm to count patterns with cycles (although does not support arbitrary patterns) and presented an MPI-based distributed implementation. However, the authors demonstrated performance on graphs with only a few million edges. 

ASAP~\cite{Iyer:2018:ASAP:222637} is a distributed solution enabling approximate match counting within a given error bound. ASAP is based on Apache Spark and GraphX~\cite{Gonzalez:2014:GGP:2685048.2685096}. Like Arabesque, ASAP provides a high-level API for implementing graph mining algorithms. ASAP implements a neighborhood sampling technique that estimates the template match count by sampling the edges in the background graph. Unlike our system, the output produced by ASAP is only probabilistic; ASAP does not offer precision and recall guarantees; although allows trade-off between the result accuracy and time-to-solution, and provides a technique to bound the counting error.

\cite{Gao.2014.6816681} introduces an approximate matching technique based on tree-search and join and evaluate it on large queries (up to 50 vertices). 
Here, a query template is converted in to a single-sink directed acyclic graph and message transition follows its topology.  Distributed approximate matching solutions based on \emph{graph simulation}~\cite{Henzinger:1995:CSF:795662.796255} are proposed in~\cite{Fard.2013.6691601} and ~\cite{Ma:2012:DGP:2187836.2187963}, although both are evaluated only on relatively small real-world graphs.


\section{Preliminaries}
\label{snx:prelm}

\noindent
We aim to identify all structures within a large \textit{background graph}, $\cG$, identical to a small connected \textit{template graph}, $\cG_0$.
We describe general graph properties for $\cG$, and use the same notation (summarized in Table~\ref{tab:notation}) for other graph objects. 

A graph $\cG(\cV, \cE)$ is a collection of $n$ vertices $\cV = \{0, 1, ..., n-1\}$ and $m$ edges $(i,j) \in \cE$, where $i,j\in \cV$ ($i$ is the edge's {\em source} and $j$ is the {\em target}).  
Here, we only discuss simple (i.e., no  self-edges), undirected, vertex-labeled graphs, although the techniques are applicable to directed, non-simple graphs, with labels on both edges and vertices.   
An {\em undirected} $\cG$ satisfies $(i,j) \in \cE$ if and only if $(j,i)\in \cE$.   
Vertex $i$'s {\em adjacency list}, $adj(i)$, is the set of all $j$ such that $(i,j) \in \cE$. 
A {\em vertex-labeled graph} also has a set of $n_\ell$ labels $\cL$ of which each vertex $i \in \cV$ has an assignment $\ell(i) \in \cL$. 


A {\em walk} 
in $\cG$ is an ordered subsequence of $\cV$ 
where each consecutive pair is an edge in $\cE$.
A walk with no repeated vertices is a {\em path}. 
A path with equal first and last vertex is a {\em cycle}. 
An {\em acyclic} graph has no cycles.


We further characterize graphs with with cycles.   Two {\em disjoint} cycles have no edge in common. Two {\em distinct} cycles have at least one edge not in common.  We define the {\em cycle degree} of edge $(i,j) \in \cE$ as the number of distinct cycles $(i,j)$ is in, written $\delta_{(i,j)}$.   The maximum cycle degree is $\delta_{max} := \max_{\cE} \delta_{(i,j)}$.   A graph is {\em edge-monocyclic} if $\delta_{max} = 1$.

We discuss several graph objects simultaneously: the {\em template graph} $\cG_0(\cV_0, \cE_0)$,  
the {\em background graph} $\cG(\cV,\cE)$, and
the {\em current solution subgraph} $\cG^*(\cV^*,\cE^*)$,
with $\cV^* \subset \cV$ and $\cE^* \subset \cE$. 
Our techniques iteratively refine $\cV^*$ and $\cE^*$ until they converge to the union of all subgraphs of $\cG$ that {\em exactly match} the template, $\cG_0$.

For clarity, when referring to vertices and edges from the template graph, $\cG_0$, we will use the notation $q_i \in \cV_0$ and $(q_i, q_j) \in \cE_0$.   
Conversely, we will use $v_i \in \cV$ and $(v_i, v_j) \in \cE$ for vertices and edges from the background graph $\cG$ or the solution subgraph $\cG^*$. In the rest of the paper, particularly in \S\ref{snx:di}, Alg.~\ref{alg:vertex_state_1},~\ref{alg:lcp_2},~\ref{alg:lcp_1},~\ref{alg:pc_2} and ~\ref{alg:pc_1},   
we use subscripts ($i$, $j$ and $k$) to differentiate between distinct vertices of the background graph $\cG$ and that of the template graph $\cG_0$. (For example, in Alg.~\ref{alg:vertex_state_1}, a vertex $v_j$'s state may be updated if it has received a message from another vertex $v_i$, where $(v_i, v_j) \in \cE$. To avoid confusion, we use a different subscript to represent a template vertex, e.g., $q_k$.
When it is clear from context, we 
adapt notation to avoid double subscripts, using $q_0$ or $v_5$ in place of $q_{i_0}$ or $v_{i_5}$. 


We assume $\cG_0$ is connected, because if $\cG_0$ has multiple components the matching problem can be easily reduced to solving it for each component individually.

\begin{definition}
\label{def:exacto}
A subgraph $\cH(\cV_\cH,\cE_\cH), \cV_\cH \subset \cV, \cE_\cH \subset \cE$ is an exact match of template graph $\cG_0(\cV_0, \cE_0)$ (in notation, $\cH \sim \cG_0$) if  there exists a bijective function
$\phi : \cV_0 \longleftrightarrow \cV_\cH$ with the properties (note that $\phi$ may not be unique for a given $\cH$):
\begin{itemize}
\item[(i)] $\ell(\phi(q)) = \ell(q)$, for all $q \in \cV_0$ and
\item[(ii)] $\forall (q_1, q_2) \in \cE_0$, we have $(\phi(q_1), \phi(q_2)) \in \cE_\cH$  
\item[(iii)] $\forall (v_1, v_2) \in \cE_\cH$, we have $(\phi^{-1}(v_1), \phi^{-1}(v_2)) \in \cE_0$
\end{itemize}
\end{definition}

\begin{table}[!t]
\footnotesize
\renewcommand{\arraystretch}{1.2}
\captionsetup{font=footnotesize}
\caption{Symbolic notation used.}
\label{tab:notation}
\centering
\begin{tabular}{l|l}
\hline
Object(s) & Notation \\ 
\hline
template graph, vertices, edges & $\cG_0 (\cV_0, \cE_0)$ \\
template graph sizes & $n_0 := |\cV_0|$, $m_0 := |\cE_0|$ \\
vertices in the template graph & $\cV_0 := \{\tvert_0, \tvert_1, ..., \tvert_{n_0-1}\}$ \\
edges in the template graph & $(\tvert_i, \tvert_j) \in \cE_0$\\
set of vertices adjacent to $q_i$ in $\cG_0$ & $adj(q_i)$\\
\hline
background graph, vertices, edges & $\cG (\cV, \cE)$ \\
background graph sizes & $n := |\cV|$, $m := |\cE|$ \\
vertices in the background graph  & $\cV := \{v_0, v_1, ..., v_{n-1}\}$\\
edges in the background graph  & $(v_i, v_j) \in \cE$\\
set of vertices adjacent to $v_i$ in $\cG$ & $adj(v_i)$\\
maximum vertex degree in $\cG$ & $d_{max}$\\
average vertex degree in $\cG$ & $d_{avg}$\\
standard deviation of vertex degree in $\cG$ & $d_{stdev}$\\
\hline
label set & $\cL = \{0, 1, ..., n_\ell-1\}$ \\
vertex label of $q_i$  & $\ell(q_i) \in \cL$ 
\\
\hline
vertex match function & $\omega(v_i) \subset \cV_0$ \\   
set of non-local constraints for $\cG_0$  & $\cK_0$\\
\hline
matching subgraph, vertices, edges & $\cH (\cV_\cH, \cE_\cH)$ \\
solution subgraph, vertices, edges & $\cG^*(\cV^*, \cE^*)$ \\
\hline
\end{tabular}
\end{table}

\textit{\textbf{Intuition for our Solution.}} The algorithms we develop here, iteratively refine a {\em vertex match} function $\omega(v) \subset \cV_0$ such that, for every $v \in \cV$, $\omega(v)$ stores a superset of all template vertices $v$ can possibly match.  Set $\omega(v)$ converges to contain all possible values of $\phi^{-1}(v)$, where $v$ is involved in one or more matching subgraphs. 
When a single constraint involving $q \in \cV_0$ is violated/unmet, $q$  is no longer a possibility for $v$ in a match and $q$ is removed: $\omega(v) \leftarrow \omega(v) \setminus \{q\}$.  


\begin{remark}
\label{rem:tds}
Given an ordered sequence of all $n_0$ vertices $\{q_1, q_2, ... , q_{n_0}\} \subset \cV_0$, a simple (although potentially expensive) search from $v_1 \in \cV^*$ verifies if $v_1$ is in a match, with $\phi(q_1) = v_1$, or not.  
The search lists an ordered sequence $\{v_1, v_2, ... , v_{n_0}\} \subset \cV^*$, with $\phi$ defined as $\phi(q_k) = v_k$.
Search step $k$ proposes a new $v_k$, checking Def.~\ref{def:exacto}~(i) and (ii).   
If all checks are passed, the search accepts $v_k$ and moves on to step $(k+1)$, but terminates if no such $v_k$ exists in $\cV^*$.   
If the full list is generated with all label and edge checks passed then there exists a $\cH \sim \cG$ with $\cV_\cH = \{v_1, v_2, ... , v_{n_0}\}$.
\end{remark}

We call this \emph{Template-Driven Search (TDS)}, presented in the next section and develop an efficient distributed version in \S\ref{snx:di}, to apply to the solution $\cG^*(\cV^*, \cE^*)$. If TDS has been applied successfully then there are no false positives remaining independently of the structure of $\cG_0$. We note that TDS is needed only for the general case, and multiple other specific cases simpler verification routines can be used.


\section{Pattern Matching via Constraint Checking -- Solution Overview}
\label{sxn:algs}

\noindent
Our goal is to realize a technique which systematically eliminates all the vertices and edges that do not participate in any exact match $\cH \sim \cG_0$. 
This approach is motivated by viewing the template $\cG_0$ as specifying a set of constraints the vertices and edges that participate in a match must meet.
As a trivial example, any vertex $v$ whose label $\ell(v)$ is not present in $\cG_0$, cannot be present in an exact match. 
A vertex in an exact match also needs to have non-eliminated edges to non-eliminated vertices labeled as prescribed in the adjacency structure of the corresponding template vertex. 

Local constraints that involve a vertex and its immediate neighborhood can be checked by having vertices communicate their `provisional' template match(es) with their one-hop neighbors in the solution subgraph $\cG^*(\cV^*,\cE^*)$ (i.e., the currently pruned background graph).  We call this process \emph{Local Constraint Checking (LCC)} and note that, since communication is limited to one-hop vertex neighbors, this is a relatively low cost step. For a restricted set of search templates, acyclic or edge-monocyclic with unique vertex labels, LLC is sufficient for a precise solution~\cite{Reza:2017:IEEECluster:PM.8048872}.  For more complex search templates, in our experimental setup, LLC  often removes the bulk of non-matching vertices and edges; although in many cases, most of the search effort is allocated to verifying the non-local constrains we describe below ~\cite{Reza:2018:SC:PM}. 

\begin{figure}[!t]
\centering
\includegraphics[width=3.5in]{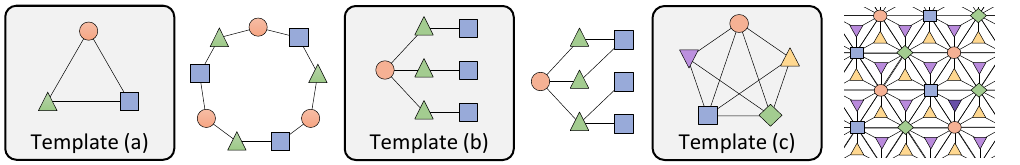}
\captionsetup{font=footnotesize}
\caption{Three examples of search templates and background graphs that justify the full set (local and non-local) of pruning constraints. Template (a) is a 3-Cycle; cycles of length 3$k$ with repeated labels in the background graph meet neighborhood constraints, surviving local constraint checking. 
Template (b) contains several vertices with non-unique labels; to its right there is a background graph that meets individual point-to-point path constraints, also surviving (non-local) path checking.   
Template (c) is characterized by two 4-Cliques that overlap at a 3-Cycle; 
the background graph structure to the right is doubly periodic (a $4 \times 3$ torus) and meets all edge and vertex cycle constraints, surviving (non-local) cycle checking. Templates (b) and (c) require template-driven search to guarantee no false positives; template (a) only needs cycle checking in addition to checking the local constraints. (Reused from~\cite{Reza:2018:SC:PM}.)
} 
\label{fig:counter_example_0012}
\end{figure}

Templates with topological requirements beyond the immediate neighborhood of a vertex (i.e., templates with cycles and/or repeated vertex labels) require additional routines to check non-local properties to guarantee that all non-matching vertices are eliminated. (Fig.~\ref{fig:counter_example_0012} illustrates the need for these additional checks with examples). To support arbitrary templates, we have developed a process which we dub \textit{Non-local Constraint Checking (NLCC)}: first, based on the search template $\cG_0$, we generate the set of constraints $\cK_0$ that are to be verified, then prune the graph using each of them.  

\subsection{Overview of the Constraint Checking Technique}
 
\begin{figure*}[!t]
\centering
\includegraphics[width=\linewidth]{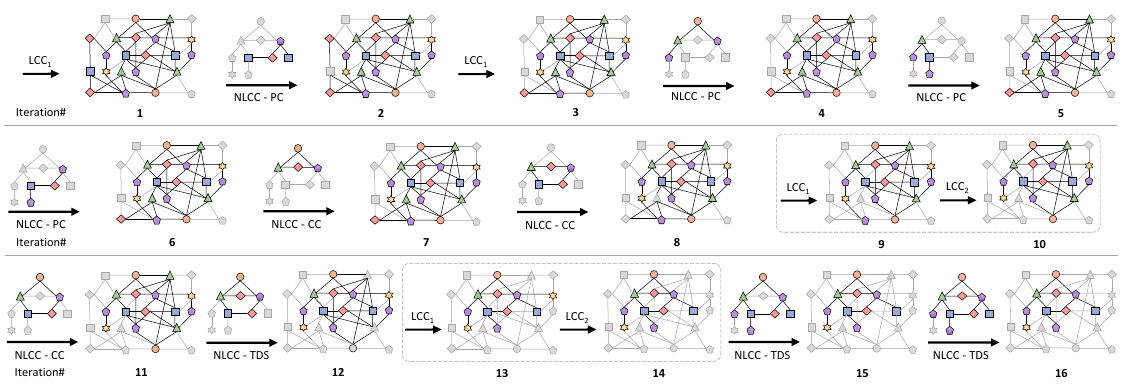} 
\captionsetup{font=footnotesize}
\caption{Algorithm walk through for the example background graph and template in Fig.~\ref{fig:example_0011}, depicting which vertices and edges in $\cG^*(\cV^*, \cE^*)$ are eliminated (in solid grey) during each iteration. 
The non-local constraints for $\cG_0$ are listed in Table~\ref{table:non-local_constraint_generation}.  The example does not show application of some of the constraints as that do not eliminate vertices or edges. (Reused from~\cite{Reza:2018:SC:PM}.)} 
\label{fig:algorithm_example_0011}
\end{figure*}

Alg.~\ref{alg:pm_1} presents an overview of our solution. This section provides high-level descriptions of the local and non-local constraint checking routines while \S\ref{snx:di} provides the detailed distributed asynchronous algorithms for a vertex-centric abstraction. As an overview, Fig.~\ref{fig:algorithm_example_0011} illustrates the complete workflow for the graph and pattern in Fig.~\ref{fig:example_0011}, for which constraint generation is detailed in Table ~\ref{table:non-local_constraint_generation}.
\\

\begin{algorithm}
\footnotesize
\caption{Main Constraint Checking Loop}
\label{alg:pm_1}
\begin{algorithmic}[1]
\State \textbf{Input:} background graph $\cG (\cV, \cE)$, template $\cG_0 (\cV_0, \cE_0)$
\State \textbf{Output:} solution subgraph $\cG^*(\cV^*,\cE^*)$ 
\State $\cK_0\leftarrow$ NON\_LOCAL\_CONSTRAINTS ($\cG_0$)
\State INIT\_VERTEX\_STATE ($\cG, \cG_0$)
\State $\cG^*\leftarrow$ LOCAL\_CONSTRAINT\_CHECKING ($\cG, \cG_0$)
\While{$\cK_0$ is not empty}   
  \State pick and remove the next constraint $\cC_0$ from $\cK_0$ 
  \State $\cG^*\leftarrow$ NON\_LOCAL\_CONSTRAINT\_CHECKING ($\cG^*, \cG_0, \cC_0$)      
  \If {any vertex has been eliminated or has one of its provisional matches removed}
  	\State $\cG^*\leftarrow$ LOCAL\_CONSTRAINT\_CHECKING ($\cG^*, \cG_0$)
  \EndIf	
\EndWhile
\State \Return $\cG^*$
\end{algorithmic}
\end{algorithm}

\textit{\textbf{Local Constraint Checking (LCC)}} 
involves a vertex and its immediate neighborhood. The algorithm performs the following two operations: \emph{(i)~Vertex elimination} - the algorithm excludes the vertices that do not have a corresponding label in the template then, iteratively, excludes the vertices that do not have neighbors as labeled in the template. For templates that have vertices with multiple neighbors with the same label, the algorithm verifies if a matching vertex in the background graph has a minimum number of distinct active neighbors with the same label as prescribed in the template. \emph{(ii)~Edge elimination} - this excludes edges to eliminated neighbors and edges to neighbors whose labels do not match the labels prescribed in the adjacency structure of its corresponding template vertex (e.g., Fig.~\ref{fig:algorithm_example_0011}, Iteration \#1). Edge elimination is crucial for scalability, since, in a distributed setting, no messages are sent over eliminated edges thus significantly improving the overall efficiency of the system (evaluated in \S\ref{sxn:evl}, Fig.~\ref{fig:strong_scaling_0011}). 
\\

\textit{\textbf{Non-local Constraint Checking (NLCC)}}
aims to exclude vertices that fail to meet topological and label constraints beyond the one-hop neighborhood, that LCC is not guaranteed to eliminate (an example is presented Fig.~\ref{fig:counter_example_0012}). We have identified three types of non-local constraints which can be verified independently: \emph{(i)} Cycle Constraints (CC), \emph{(ii)} Path Constraints (PC), and \emph{(iii)} constraints that require Template-Driven Search (TDS) (see Remark~\ref{rem:tds}). For arbitrary templates, TDS constraints based on aggregating multiple paths/cycles enable further pruning, and 
insure that pruning yields no false positives. Compared to CC and PC, checking TDS constraints, however, can be more expensive. To reduce the overall cost, we first generate single cycle- and path-based constraints, which are usually less costly to verify, and prune the graph using them before deploying TDS (the effectiveness of this ordering is evaluated in Fig.~\ref{fig:perf_comp_0011}(c)).
\\

\begin{table}[!t]
\footnotesize
\setlength{\tabcolsep}{2.0pt}
\renewcommand{\arraystretch}{1.5}
\captionsetup{font=footnotesize}
\caption{Step-by-step illustration of non-local constraint generation for the template in Fig.~\ref{fig:example_0011}; high-level pseudocode, accompanied by pictorial depiction. The figures show the steps to generate the required cycle constraints (CC), path constraints (PC), and higher-order constraints requiring template-drive search (TDS). Alg.~\ref{alg:tds_generation} is the pseudocode for the procedure TDS\_CONSTRAINTS(). Definitions of the helper functions UNIQUE\_LABEL\_VERTICES(), LEAF\_VERTICES() and FIND\_CIRCLES() are rather trivial; hence, not included. (Figures adapted from~\cite{Reza:2018:SC:PM}.)
}
\label{table:non-local_constraint_generation}
\centering
\begin{tabular}{c|m{8cm}|>{\centering\arraybackslash}m{4.8cm}}
\hline
{\rotatebox[origin=m]{90}{}} &  

\begin{algorithmic}[1]
\State \textbf{Input:} template $\cG_0 (\cV_0, \cE_0)$ \par
\Comment{$\cG_0$ is undirected and at least weakly connected}
\State \textbf{Output:} non-local constraint set $\cK_0$ of $\cG_0$ 
\Procedure{non\_local\_constraints}{$\cG_0$}
\State $\cK_0 \leftarrow \emptyset$;
\quad $CC \leftarrow \emptyset$;
\quad $PC \leftarrow \emptyset$; \par
\hspace{-0.5em}
$TDS_{CC} \leftarrow \emptyset$;
\quad $TDS_{PC} \leftarrow \emptyset$;
\quad $TDS \leftarrow \emptyset$
\par\Comment{$\cK_0$ is an ordered set; the rest are lists for each constraint type}
\algstore{bkbreak}
\end{algorithmic} 

& 

\includegraphics[width=1.8in]{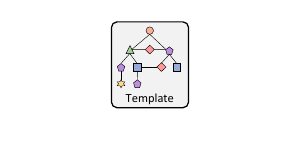}
\\
\hline
{\rotatebox[origin=m]{90}{ Vertex Classification }} & 

\begin{algorithmic}[1]
\algrestore{bkbreak}
\State $\cV_u \leftarrow \text{UNIQUE\_LABEL\_VERTICES}(\cV_0\in\cG_0)$
\State $\cV_{ul} \leftarrow \text{LEAF\_VERTICES}(\cV_u\in\cG_0)$ 
\Comment{Step 1}
\State $\cV'_0 \leftarrow \cV_0\setminus\cV_{ul}$
\State $\cE'_0 \leftarrow \forall(q_i,q_j)\in\cE_0 \textbf{ where } q_i\in\cV'_0 \textbf{ and } q_j\in\cV'_0$
\State $\cG'_0 \leftarrow (\cV'_0, \cE'_0)$
\State $\cV_{d} \leftarrow \cV_0\setminus\cV_{u}$
\Comment{Step 2}
\algstore{bkbreak}
\end{algorithmic}

& 

\includegraphics[width=1.8in]{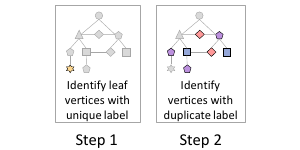}
\\
\hline
{\rotatebox[origin=m]{90}{ Cycle Constraints }} &


\begin{algorithmic}[1]
\algrestore{bkbreak}
\State $CC \leftarrow \text{FIND\_CIRCLES}(\cG'_0)$ 
\algstore{bkbreak}
\end{algorithmic}

&

\includegraphics[width=1.8in]{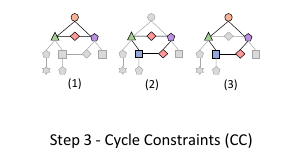}
\\
\hline
{\rotatebox[origin=m]{90}{Path Constraints}} & 


\begin{algorithmic}[1]
\algrestore{bkbreak}
\ForAll{ vertex pairs $\{q_i,q_j\} \subset \cV_d \textbf{ where } \ell(q_i)=\ell(q_j)$}
\State $path \leftarrow \text{FIND\_SHORTEST\_PATH}(\cG_0,q_i,q_j,3)$ \par
\Comment{a unique shortest path from $q_i \text{ to } q_j$ of length $\geq$ 3 }
\If{$path \neq \emptyset$ }
$PC.add(path)$ 
\EndIf
\EndFor
\algstore{bkbreak}
\end{algorithmic}

&

\includegraphics[width=1.8in]{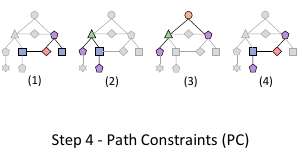}
\\
\hline
{\rotatebox[origin=m]{90}{TDS Constraints}} &




\begin{algorithmic}[1]
\algrestore{bkbreak}
\State $TDS_{CC} \leftarrow \text{TDS\_CONSTRAINTS}(CC) $
\par\Comment{TDS cycle constraints, Step 5(1)}
\State $TDS_{PC} \leftarrow \text{TDS\_CONSTRAINTS}(PC) $
\par\Comment{TDS path constraints, Step 5(2)}
\State $TDS \leftarrow TDS_{CC} \cup TDS_{PC}$
\State $TDS \leftarrow \text{TDS\_CONSTRAINTS}(TDS) $
\par\Comment{TDS constraints, Step 5(3)}
\algstore{bkbreak}
\end{algorithmic}

&

\includegraphics[width=1.8in]{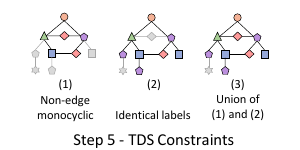}
\\
\hline
{\rotatebox[origin=m]{90}{}} &

\begin{algorithmic}[1]
\algrestore{bkbreak}
\State \Return $\cK_0 \leftarrow PC \cup CC \cup TDS_{CC} \cup TDS_{PC} \cup TDS$ \par  
\EndProcedure
\end{algorithmic}

&
\\
\hline
\end{tabular}
\end{table}

\textit{High-level Algorithmic Approach.} Regardless of the constraint type, NLCC leverages a \emph{token passing} approach: tokens are issued by background graph vertices whose corresponding template vertices are identified to have non-local constraints.
After a fixed number of steps, we check if a token has arrived where expected (e.g., back to the originating vertex for checking the existence of a cycle). If not, then the token issuing vertex does not satisfy the required constraint and is eliminated. Along the token path, the algorithm verifies that all expected labels are encountered and, where necessary, uses the path information accumulated with the token to verify that different/repeated vertex identity constraint expectations are met. Next, we discuss how each type of non-local constraint is verified.
\\

\textit{Cycle Constraints (CC).} Higher-order structures within $\cG$ that survive LCC, but do not contain $\cG_0$, are possible if $\cG_0$ contains a cycle (this happens if $\cG$ contains one or more \emph{unrolled} cycles as in Fig.~\ref{fig:counter_example_0012}, Template (a)). To address this, we directly check for cycles of the correct length. 
\\

\textit{Path Constraints (PC).} If the template $\cG_0$ has two or more vertices with the same label that are three or more hops away from each other, then structures in $\cG$ that survive LCC, yet contain no match, are possible (Fig.~\ref{fig:counter_example_0012}, Template (b)). Thus, for every vertex pair with the same label in $\cG_0$, we directly check the existence of a path of the correct length and label sequence for prospective matching vertices in $\cG^*$. Opposite to cycle checking, after a fixed number of steps, a token must be received by a vertex {\em different} from the token initiating vertex but with an identical label. 
\\

\textit{TDS Constraints.} 
These are partial (for further pruning and performance optimization similar to path- and cycle-constraints) 
or complete (i.e., including all edges of the template) walks on the template (required to ensure correctness). The token \emph{walks} the constraint in the background graph and verifies that each vertex visited meets its neighborhood constraints (Remark~\ref{rem:tds}). In a distributed memory setting, this is done by maintaining a history of the walk and checking that previously visited vertices are revisited as expected.  
TDS constraints are crucial to guarantee zero false positives for templates that are \emph{non-edge-monocyclic} or have repeated labels (Fig.~\ref{fig:counter_example_0012}, Template (b) and (c)). Next, we describe how these three types of non-local-constraints are generated.

\begin{algorithm}
\footnotesize
\caption{TDS Constraint Generation}
\label{alg:tds_generation}
\begin{algorithmic}[1]
\Procedure{tds\_constraints}{$\cK'_0$}
\State $i \leftarrow 0$; \quad$j \leftarrow 1$
\ForAll{$\cK'_0[i] \in \cK'_0; \quad i \leftarrow i+1$}
\ForAll{$\cK'_0[j] \in \cK'_0; \quad j \leftarrow j+1$}

\If{$\cK'_0[i] \textbf{ and } \cK'_0[j] \text{ have at least one common vertex}$}

\If{$\cK'_0[i] \textbf{ and } \cK'_0[j] \text{ have at least one common edge} \textbf{ or } type(\cK'_0)=PC $}

\State $\cK'_0[i] \leftarrow \cK'_0[i] \cup \cK'_0[j]$
\Comment{union of two constraints (i.e., substructures of $\cG_0$)}

\State $\cK'_0 \leftarrow \cK'_0 \setminus \cK'_0[j]$;
\quad $j \leftarrow j-1$

\EndIf
\EndIf

\EndFor
\EndFor

\If{$type(\cK'_0)=PC$}
\ForAll{$\cC_0 \in \cK'_0$}
\State $\cC_0 \leftarrow \text{SPANNING\_TREE}(\cC_0)$
\EndFor
\EndIf

\State \Return $\cK'_0$

\EndProcedure

\end{algorithmic}
\end{algorithm}

\subsection{Non-local Constraint Generation}
We generate non-local constraints following the heuristic presented in Table~\ref{table:non-local_constraint_generation}. The three types of non-local constraints, namely, Cycle Constraints, Path Constraints and TDS Constraints are generated incrementally: for an example template, we provide a step-by-step illustration of the non-local constraint generation. Fig.~\ref{fig:algorithm_example_0011} shows a complete example of how pruning progresses using the generated constraints.  

\textit{Step 1.} Identify all the leaf vertices (i.e., a vertex with only one neighbor) with unique labels. They are not considered for non-local constraint checking as LCC guarantees pruning if there is no match. 

\textit{Step 2.} Identify all the vertices with duplicate labels. Path constraints are generated only for these vertices.

\textit{Step 3.} If the template has cycles, then individual cycles are identified and a cycle constraint is generated for each cycle.

\textit{Step 4.} 
For all possible combinations of vertex pairs with identical label, we identify all existing paths greater than or equal to three-hop length. (LCC precisely checks identical label pairs that are one or two hops from each other). One such path, for each vertex pair, is generated as a path constraint. Here, two optimization's are applied to minimize the number of path constraints to be verified: \textit{(i)} If there are multiple paths connecting two terminal vertices then the shortest path is generated as a path constraint. \textit{(ii)} If all the edges comprising a path also belong to a cycle constraint, that particular path is excluded from the set of path constraints. Verification of the cycle constraint will implicitly check for existence of a successful walk of appropriate length connecting the terminal vertices (of the path of interest). 

\textit{Step 5.} We generate TDS constraints in three steps. First, for templates with multiple cycles sharing more than one vertex (i.e., the template is non-edge-monocyclic), a TDS cyclic constraint is generated through the union of previously identified cycle constraints. This results in a higher-order cyclic structure with a maximal set of edges that cover all the cycles sharing at least one edge (e.g., Step 5(1)).

Second, for templates with repeated labels, a new TDS constraint is generated through the union of all previously identified path constraints. This procedure generates higher-order structure that covers all the template vertices with repeated labels (e.g., Step 5(2)).

The final step generates a TDS constraint as the union of the previously identified two constraints (e.g., Step 5(3)). Note that the above is a heuristic, more TDS constraints can be generated by creating various possible combinations of cycles and paths. Only this third step is mandatory to eliminate all false positives.
\\

\textit{\textbf{Constraint Optimization.}} 
Non-local constraint verification checks for existence of at least one successful walk of the appropriate length. There are alternatives to how tokens could be passed around to complete a walk. The non-local constraint generation also focuses on optimizing the walks for token passing.  

Whenever possible, we orchestrate each walk so the vertices are visited in the increasing order of label frequency in the background graph. (This procedure has negligible overhead as label frequency is computed only once per label set and we only sort the vertex list of a template which, typically, has $10^1$--$10^2$ elements). Here, the goal is to curb combinatorial growth of the algorithm state (or more specifically, in the distributed memory setting, the number of messages). This optimization has the potential of eliminating a large part of the graph without explorations deep into an excessive number of branches in the backgorund graph.

Non-local constraint generation also focuses of reducing the number of constraints and length of a walk (as the size of a constraint directly influences complexity). In addition to selecting the shortest path, if all the edges in a path constraint are also present in a cycle constraint, the path is ignored. When generating TDS constraints through union of the original path constrains, we are able to remove some (redundant) edges by obtaining a spanning tree for each TDS constraint (Alg.~\ref{alg:tds_generation}, line \#11). 
\\

\textit{Constraint Ordering Heuristics.} 
We use a second set of heuristics to optimize the order in which constraints are scheduled for verification. First, we check for path and cycle constraints, since they tend to be less expensive than TDS constraints. Second, we order the non-local constraints with respect to increasing length of the walk as longer walks are more susceptible to combinatorial explosion.  ~\cite{Nicolas:2018:IA3:PM:8638389} presents an avenue to design advanced heuristics.
\\

\textit{Token Generation.} 
For cyclic constraints, a token must be initiated from each vertex that may participate in the substructure, whereas for path constraints, tokens are only initiated from terminal vertices. Tokens are started from vertices (that belong to the same cyclic substructure) in the increasing order of their label frequency in the background graph. For duplicate/distinct label verification, there is also TDS path constraint checking. The substructure in question may contain a cycle or a tree. Similar to path constraints, here, tokens are initiated from vertices with duplicate labels.


\begin{algorithm}
\footnotesize
\caption{Vertex State and Initialization}
\label{alg:vertex_state_1}
\begin{algorithmic}[1]

\State vertex state: $\alpha(v_j) \leftarrow false$
\Comment{indicates if a vertex $v_j$ is active ($true$) or eliminated ($false$)}
\State vertex state: $\omega(v_j) \leftarrow \emptyset$
\Comment{list of possible vertex matches in template}
\State vertex state: $\varepsilon(v_j)$
\Comment{map of active edges of a vertex $v_j$}
\State vertex state: $\tau(v_j) \leftarrow \emptyset$
\Comment{set of already forwarded tokens by vertex $v_j$, used for work aggregation in NLCC (Alg.~\ref{alg:lcp_2})}

\Procedure{init\_vertex\_state}{$\cG, \cG_0$}
\ForAll{$v_j\in\cV$}
\ForAll{$q_k\in\cV_0$}
\If{$\ell(q_k)=\ell(v_j)$}
\If{$\alpha(v_j)=false$}
\State $\alpha(v_j) \leftarrow true$
\EndIf
\State $\omega(v_j).add(q_k)$ 
\EndIf
\EndFor
\If{$\alpha(v_j)=true$}
\ForAll{$v_i \in adj(v_j)$}
\State $\varepsilon(v_j).insert(v_i, \emptyset)$ 
\EndFor
\Else{}
\State $\varepsilon(v_j) \leftarrow \emptyset$ 
\Comment{eliminate edges of an inactive vertex}
\EndIf
\EndFor
\EndProcedure

\end{algorithmic}
\end{algorithm}

\section {Distributed System Design and Implementation}
\label{snx:di}

\noindent
In this section, we present the constraint checking algorithms in the \emph{vertex-centric} abstraction of HavoqGT \cite{havoqgt.001}, an MPI-based framework that supports \emph{asynchronous} graph algorithms in the distributed environment. Our choice for HavoqGT is driven by multiple considerations: First, unlike most graph processing frameworks that primarily support the Bulk Synchronous Parallel (BSP) model, HavoqGT has been designed to support asynchronous algorithms which is essential to achieve high-performance. Asynchronous algorithms can exploit the low latency ($\sim$1$\mu$s) interconnect on leadership-class High Performance Computing (HPC) platforms. Second, the framework has demonstrated excellent scaling properties for a number of graph traversal problems~\cite{Pearce:2013:STM:2510661.2511411, Pearce:2014:FPT:2683593.2683654}. Finally, it enables load balancing: HavoqGT's \emph{delegate partitioned graph} distributes the edges of each high-degree vertex across multiple compute nodes, which is crucial for achieving scalability for scale-free graphs with skewed degree distribution. 

In HavoqGT, graph algorithms are implemented as vertex-callbacks: the user-defined $visit()$ callback can only access and update the state of a vertex.  
The framework offers the ability to generate events (a.k.a. \emph{visitors} in HavoqGT's vocabulary) that trigger this callback - either at the entire graph level using the $do\_traversal()$ method, or for a neighboring vertex using the $push(visitor)$ call. When a vertex wants to pass data to a neighbor, invoking $push(visitor)$ enqueues the relevant visitor to the distributed message queue, which exploits MPI asynchronous communication primitives for exchanging messages. This enables asynchronous vertex-to-vertex communication. 
The asynchronous graph computation completes when all events have been processed, which is determined by a distributed quiescence detection algorithm~\cite{Wellman.2000.858469}.

Alg.~\ref{alg:pm_1} outlines the key steps of the graph pruning procedure. Below, we describe the distributed implementation of the local and non-local constraint checking, and match enumeration routines. Alg.~\ref{alg:vertex_state_1} lists the state maintained by each active vertex and its initialization. 

\begin{algorithm}
\footnotesize
\caption{Local Constraint Checking}
\label{alg:lcp_2}
\begin{algorithmic}[1]
\State $\eta(v_s, v_j)$ - verifies if $v_s$ satisfies a local constraint of $v_j$; returns $\omega(v_s)$ if constraints are met, $\emptyset$ otherwise

\Procedure{local\_constraint\_checking($\cG, \cG_0$)}{}
	\Do    	
    	\State $do\_traversal(msg_{type}\leftarrow init)$ 
    	\State \textbf{barrier}
		\ForAll {$v_j\in\cV$}
        	\State $\omega^\prime\leftarrow\emptyset$ \Comment set of template matches for neighbors of $v_j$
        	\ForAll {$v_i\in\varepsilon(v_j)$}
                \If {$\eta(v_i, v_j) = \emptyset$}
                    \State $\varepsilon(v_j).remove(v_i)$ 
                    \Comment edge eliminated
                    \State \textbf{continue}
                \Else 
                \State $\omega^\prime\leftarrow\omega^\prime\cup\eta(v_i, v_j)$
                \Comment accumulate matched neighbor information
                \State reset the value field of $v_i \in \varepsilon(v_j)$ for the next iteration
                \EndIf     
        	\EndFor
            \ForAll {$q_k \in \omega(v_j)$} 
            \Comment for each potential match
                \If {$adj(q_k)\nsubseteq\omega^\prime$}  
                \State $\triangleright$ $q_k$ does not meet neighbor requirements
                    \State $\omega(v_j).remove(q_k)$ 
                    \Comment remove from the set of potential matches
                    \State \textbf{continue}
                \EndIf
            \EndFor
            \If {$\varepsilon(v_j) = \emptyset$ \textbf{or} $\omega(v_j) = \emptyset$}
                \State $\alpha(v_j)\leftarrow false$
                \Comment vertex eliminated 
            \EndIf	
            \EndFor
	\doWhile vertices or edges are eliminated
    \Comment global detection
\EndProcedure
\end{algorithmic}
\end{algorithm}

\begin{algorithm}
\footnotesize
\caption{Local Constraint Checking Visitor}
\label{alg:lcp_1}
\begin{algorithmic}[1]
\State visitor state: $v_j$ - vertex that is visited
\State visitor state: $v_s$ - vertex that originated the visitor
\State visitor state: $\omega(v_s)$ - set of possible matches in template for vertex $v_s$ 
\State visitor state: $msg_{type}$ - $init$ or $alive$
\Procedure{visit}{$\cG, vq$}
    \Comment {$vq$ - visitor queue (the distributed message queue)}
    \If {$\alpha(v_j)=false$}
    	\Return     
    \EndIf    
    \If {$msg_{type}=init$}
      \ForAll{$v_i \in \varepsilon(v_j)$}
          \State $vis$ $\leftarrow$ LCC\_VISITOR($v_i$, $v_j$, $\omega(v_j)$, $alive$)
          \State $vq.push(vis)$         
      \EndFor
    \ElsIf {$msg_{type}=alive$}
        \State $\varepsilon(v_j).get(v_s) \leftarrow \omega(v_s)$
    \EndIf
\EndProcedure
\end{algorithmic}
\end{algorithm}

\subsection{Local Constraint Checking} 
Local Constraint Checking is implemented as an iterative process (Alg.~\ref{alg:lcp_2} and the corresponding callback, Alg.~\ref{alg:lcp_1}). Each iteration initiates an asynchronous traversal by invoking the $do\_traversal()$ method and, as a result, each active vertex receives a visitor with $msg_{type} = init$. In the triggered $visit()$ callback, if the label of a vertex $v_j$ in the graph is a match for the label of any vertex in the template and the vertex is still active, it creates visitors for all its active neighbors in $\varepsilon(v_j)$ with $msg_{type} = alive$  (Alg.~\ref{alg:lcp_1}, line \#9). When a vertex $v_j$ is visited with $msg_{type} = alive$, it verifies whether the sender vertex $v_s$ satisfies one of its own (i.e., $v_j$'s) local constraints by invoking the function $\eta(v_s, v_j)$. By the end of an iteration, if $v_j$ satisfies all the template constraints, i.e, it has neighbors with the required labels (and, if needed, a minimum number of distinct neighbors with the same label as prescribed in the template), it stays active (i.e., $\alpha(v_j) = true$) for the next iteration. For templates that have multiple vertices with the same label, in any iteration, a vertex with that label in the background graph could match any of these vertices in the template, so each match must be verified independently. If $v_j$ fails to satisfy the required local constraints for a template vertex $q_k \in \omega(v_j)$, $q_k$ is removed from $\omega(v_j)$. At any stage, if $\omega(v_j)$ becomes empty, then $v_j$ is marked inactive ($\alpha(v_j) \leftarrow false$) and never communicate with its neighbors again. Edge elimination excludes two categories of edges: first, the edges to neighbors, $v_i \in \varepsilon(v_j)$ from which $v_j$ did not receive a message of type $alive$, and, second, the edges to neighbors whose labels do not match the labels prescribed in the adjacency structure of the corresponding template vertex/vertices in $\omega(v_j)$. A vertex $v_j$ is also marked inactive if its active edge list $\varepsilon(v_j)$ becomes empty. Iterations continue until no vertex or edge is marked inactive.

\subsection{Non-local Constraint Checking} 
Non-local Constraint Checking iterates over $\cK_0$, the set of non-local constraints to be checked, and validates each $\cC_0\in\cK_0$ one at a time. Alg.~\ref{alg:pc_2} describes the solution to verify a single constraint: tokens are initiated through an asynchronous traversal by invoking the $do\_traversal()$ method and, as a result, each active vertex receives a visitor with $msg_{type} = init$. Each active vertex $v_j\in\cG^*$ that is a potential match for the template vertex $q_0$ at the head of a walk (i.e., a non-local constraint) $\cC_0$, broadcasts a token to all its active neighbors in $\varepsilon(v_j)$ with $msg_{type} = forward$. A map $\gamma$ is used to track these token issuers. A $token$ is a tuple $(t,r)$ where $t$ is an ordered list of vertices that have forwarded the token and $r$ is the hop counter; $t_0 \in t$ is the token-issuing vertex in $\cG^*$. The ordered list $t$ is essential for TDS since it enables detection of distinct vertices with the same label in the token path. For simpler templates, such as templates with unique vertex labels and only edge-monocycles, $t$ may only contain $t_0$ to keep the message size small.

When an active vertex $v_j$ receives a token with $msg_{type} = forward$, it verifies that if $\omega(v_j)$ is a match for the next entry in $\cC_0$, if it has received the token from a valid neighbor (with respect to entries in $\cC_0$), and that the current hop count is less than $|\cC_0|$. If these requirements are satisfied (i.e., $\mu(v_j,\cC_0,token)$  returns $true$), $v_j$ sets itself as the forwarding vertex ($v_j$ is added to $t$), increments the hop count, and broadcasts the token to all its active neighbors in $\varepsilon(v_j)$. If any of the constraints are not met, $v_j$ drops the token. If the hop count $r$ is equal to $|\cC_0|$ and $v_j$ is the same as the source vertex in the token, for a cyclic template, a cycle has been found and $v_j$ is marked $true$ in $\gamma$. For path constraints, an acknowledgement is sent to the token issuer to update its status in $\gamma$ (Alg.~\ref{alg:pc_1}, lines \#28 -- \#31). Once verification of a constraint $\cC_0$ has been completed, the vertices that are not marked $true$ in $\gamma$, are invalidated/eliminated, i.e., $\alpha(v_j) \leftarrow false$ (Alg. \ref{alg:pc_2}, line \#9).
\\

Our distributed implementation incorporates a number of design features aimed at improving performance, scalability, robustness and efficiency; we offer, a light-weight yet highly effective technique, called \emph{work aggregation}, to prevent relaying duplicate messages; and the ability to load balance an intermediate pruned graph. In the remaining of the section, we first discuss these optimizations; we then provide details about how vertex metadata (labels) are managed, and various results our system can output.

\begin{algorithm}
\footnotesize
\caption{Non-local Constraint Checking}
\label{alg:pc_2}
\begin{algorithmic}[1]


\Procedure{non\_local\_constraint\_checking($\cG, \cG_0, \cC_0$)}{}
    	\State$\gamma \leftarrow$ map of token source vertices (in $\cG$) for $\cC_0$; the value field (initialized to \textit{false}) is set to \textit{true} if the token source \indent vertex meets the requirements of $\cC_0$     	
    	\State $do\_traversal(msg_{type}\leftarrow init)$ 
    	\State \textbf{barrier}               
    	\ForAll {$v_j \in \gamma$}
        	\If {$\gamma.get(v_j) \neq true$}         	         
        	    \State $\omega(v_j).remove(q_0)$ \textbf{where} $q_0$ is the first vertex in $\cC_0$ 
        	    \Comment{violates $\cC_0$, eliminate potential match}
        	    \If {$\omega(v_j) = \emptyset$} \Comment no potential match left 
            	    \State $\alpha(v_j) \leftarrow false$ \Comment vertex eliminated
            	\EndIf    
            \EndIf
    	\EndFor
    	\State $\forall v_j \in \cV$, reset $\tau(v_j)$
\EndProcedure
\end{algorithmic}
\end{algorithm}

\begin{algorithm}
\footnotesize
\caption{Non-local Constraint Checking Visitor}
\label{alg:pc_1}
\begin{algorithmic}[1]
\State visitor state: $v_j$ - vertex that is visited 
\State visitor state: $token$ - the token is a tuple $(t,r)$ where $t$ is an ordered list of vertices that have forwarded the token and $r$ is the hop counter; $t_0 \in t$ is the vertex that originated the token
\State visitor state: $msg_{type}$ - $init$, $forward$ or $ack$
\State $\mu(v_j,\cC_0,token)$ - verifies if $v_j$ satisfies requirements of $\cC_0$ for the current state of $token$; returns $true$ if constraints are met, $false$ otherwise


\Procedure{visit}{$\cG, vq$}
    \If {$\alpha(v_j)=false$}
    	\Return 
    \EndIf       
    \If {$msg_{type}=init$ \textbf{and} $\exists q_k \in \omega(v_j)$ \textbf{where} $q_k = q_0 \in \cC_0$}
      \State $\triangleright$ initiate a token; {$v_j$ is the token source}
      \State $t.add(v_j)$;
      \quad$r \leftarrow 1$;
      \quad$token \leftarrow (t,r)$;
      \quad$\gamma.insert(v_j, false)$
      \ForAll{$v_i \in \varepsilon(v_j)$}
       	\State $vis$ $\leftarrow$ NLCC\_VISITOR($v_i, token, forward$) 
          \State $vq.push(vis)$       
      \EndFor
    \ElsIf {$msg_{type}=forward$} \Comment{$v_j$ received a token} 
        \If{$token \notin \tau(v_j)$} \Comment work aggregation optimization
            \State $\tau(v_j).insert(token)$
        \Else {}
            \Return 
            \Comment{ignore if $v_j$ already forwarded a copy of $token$}
        \EndIf  
        \If {$\mu(v_j,\cC_0,token) = true$ \textbf{and} $token.r<|\cC_0|$}     
        \State $\triangleright$ the walk can be extended with $v_j$ and it has not reached the length $|\cC_0|$ yet
            \State $token.t.add(v_j)$;\quad
            $token.r \leftarrow token.r + 1$;\quad
        	\ForAll{$v_i \in \varepsilon(v_j)$} \Comment forward the token
          	    \State $vis$ $\leftarrow$ 
          	      NLCC\_VISITOR($v_i, token, forward$)
          	    \State $vq.push(vis)$   
            \EndFor
    	\ElsIf {$\mu(v_j,\cC_0, token) = true$ \textbf{and} $token.r=|\cC_0|$}
    	    \State $\triangleright$ the walk has reached the length $|\cC_0|$ 
    	    \If{$\cC_0$ is cyclic \textbf{and} $t_0 = v_j$}
                \State{$\gamma.get(v_j) \leftarrow true$} \Return 
                \Comment{$v_j$ meets requirements of $\cC_0$}
            \ElsIf{$\cC_0$ is acyclic \textbf{and} $t_0 \ne v_j$}
                \State $vis$ $\leftarrow$ NLCC\_VISITOR($t_0, token, ack$)
          	    \State $vq.push(vis)$ 
          	    \Comment send $ack$ to the token originator, $t_0 \in t$ 
            \EndIf 
        \EndIf
    \ElsIf {$msg_{type}=ack$}    
        \State $\gamma.get(v_j) \leftarrow true$
        \Return 
        \Comment {$v_j$ meets requirements of $\cC_0$}
	\EndIf	
\EndProcedure
\end{algorithmic}
\end{algorithm}

\subsection{Work Aggregation}
All NLCC constraints attempt to identify if a walk exists from a vertex with a fixed label and through vertices with specific labels. Since the goal is to identify the existence of any such path and multiple intermediate/complete paths in the background graph often exist, to prevent combinatorial explosion, our duplicate work detection mechanism prevents an intermediary vertex (in the token path) from forwarding a duplicate token. 
NLCC uses an unordered set $\tau(v_j)$ (Alg.~\ref{alg:vertex_state_1}, line \#4) for work aggregation (see Alg.~\ref{alg:pc_1}, line \#14): at each vertex, this is used to detect if another copy of a $token$ has already visited the vertex $v_j$ taking a different path. The performance impact of this optimization is evaluated in \S\ref{snx:optimization_evaluation}. 

\subsection{Load Balancing}
Load imbalance issues are inherent to problems involving irregular data structures, such as graphs, especially when these need to be partitioned for processing over multiple nodes. For our pattern matching solution, load imbalance can be further caused by two artifacts: First, over the course of execution our solution causes the workload to mutate, i.e., we prune away vertices and edges. Second, the distribution of matches in the background graph may be nonuniform: the vertices and edges that participate in matches, may reside on a small, potentially concentrated, part of the graph. (In \S\ref{sxn:load_balancing_evaluation}, we present a detailed characterization of these artifacts.)

The iterative nature of the constraint checking pipeline allows us to adopt a \emph{pseudo-dynamic} load balancing approach: First, we checkpoint the current state of execution (at the end of an asynchronous constraint checking phase): the pruned graph, i.e., the set of active vertices and edges and the per-vertex state indicating template matches, $\omega(v_j)$ (Alg.~\ref{alg:vertex_state_1}). Next, using HavoqGT's distributed graph partitioning module, we reshuffle the vertex-to-processor assignment to evenly distribute vertices (with $\omega(v_j)$ remained intact) and edges across processing cores. Processing is then resumed on the rebalanced workload. Furthermore, depending on the size the the pruned graph, it is possible to resume processing on a smaller deployment (primarily for efficiency reasons, such as conserving CPU Hours). Over the course of the execution, checkpointing and rebalancing can be repeated as needed. We evaluate the effectiveness of different load balancing strategies and present an analysis of their impact on performance in \S\ref{sxn:load_balancing_evaluation}.

\subsection{Termination and Output}
If NLCC is not required, the search terminates when no vertex is eliminated (or none of its provisional matches is removed) in an LCC iteration. Otherwise, the search terminates when all constraints in $\cK_0$ have been verified. 
The output of constraint checking is: \textit{(i)} the set of vertices and edges that survived the iterative elimination process and, \textit{(ii)} for each vertex in this set, the mapping in the template where a match has been identified.

A distributed match enumeration or counting routine can operate on the pruned solution subgraph: 
Alg.~\ref{alg:pc_1} can be slightly modified to obtain the enumeration of the matches in the background graph; here, the constraint used is a walk on the full template, work aggregation is turned off, and each possible match is verified. 
For each of the vertices that remains in the solution set, the pruning procedure collects their exact match(es) to the search template. We use this information to accelerate match enumeration.

\subsection{Metadata Store}
The metadata is stored independent of the graph topology itself which uses the Compressed Sparse Row (CSR) format \cite{Bell:2009:ISM:1654059.1654078}. 
At initialization, only the required attributes are read from the file(s) stored on a distributed file system. A light-weight distributed process builds the in-memory (or memory-mapped) metadata store. For example, On 256 compute nodes, for the 257 billion edge Web Data Commons graph \cite{wdc.2012}, the metadata store can be populated in under a minute. Although, in this work, we consider vertex metadata (i.e., labels) only, support for edge metadata is trivial within the presented infrastructure.


%
%
\section{Complexity Analysis}
\label{sxn:complexity_analysis}

We attempt to estimate the space, time and generated message 
complexity for both local constraint checking (LCC) and non-local constraint checking (NLCC) routines presented in \S\ref{snx:di}. Note that except for the first iteration of LCC, constraint checking routines are invoked on the current (pruned) solution subgraph $\cG^*(\cV^*,\cE^*)$ where $|\cG^*|\leq|\cG|$. (See Table~\ref{tab:notation} for the symbolic notation used in this section.) 

\subsection{Local Constraint Checking}
\label{sxn:ca_lcc}
We mainly focus on analyzing the complexity of one iteration of the LCC routine presented in Alg.~\ref{alg:lcp_2}.
\\

\textit{\textbf{Space Complexity.}}
In each iteration of LCC, each active vertex $v_i \in \cV^*$ maintains a set of its template vertex matches/exclusions $\omega(v_i)$ where $|\omega(v_i)| = |\cV_0|$. Therefore, space complexity of LCC is linear in the size of the template: $O(|\cV^*|\times|\cV_0|)$. In our implementation, we use a bit vector to store the template vertex matches to reduce memory overhead. For example, if the template has 64 vertices, per-vertex (of $\cG^*$) storage requirement is eight bytes. Additionally, in one iteration of LCC, an active vertex creates one visitor per active edge, therefore, the storage requirement for the visitor queue (the message queue in HavoqGT) is $O(|\cE^*|)$.
\\

\textit{\textbf{Time Complexity.}} 
In each iteration of LCC, all active vertices in $\cV^*$ visit all their respective active neighbors (in $\cE^*$). In iteration $k$, only the vertices and edges that survived iteration $k-1$, are considered. Therefore, the time complexity of the $k$-th iteration is $O(|\cV_{k-1}^*|+|\cE_{k-1}^*|)$. Initially, i.e., when $k=0$ and no vertices and edges have been eliminated, i.e., $\cV^*=\cV$ and $\cE^*=\cE$; we can write time complexity of the first iteration is $O(|\cV^*|+|\cE^*|)$, the most expensive of all LCC iterations. Assume LCC stops eliminating vertices and edges after $k_{max}$ iterations; hence, total time complexity of LCC is $O(k_{max}\times(|\cV^*|+|\cE^*|))$. For an acyclic template with unique labels, $k_{max} = \mbox{diam}\left(\cG_0\right) + 1$ (see~\cite{Reza:2017:IEEECluster:PM.8048872}  
for proof). An analysis for the worst case for an arbitrary template does not take us far - the upper bound of maximum number of iteration in LCC is $k_{max}\leq|\cE|$. In practice, the worst case is when in each iteration only a few or no vertices and/or edges are eliminated and a large number of iterations is needed. However, for real-world, scale-free graphs, the first few steps of LCC reduce $|\cG|$ by several orders of magnitude, yielding costs nowhere near the worst case bounds (see the evaluation section (\S\ref{sxn:evl}) for multiple examples).
\\

\textit{\textbf{Message Complexity.}} 
In each iteration, an active vertex creates one visitor per active edge, resulting in one message per edge. The analysis is similar to the one above: the message complexity of one iteration of LCC is $O(|\cE^*|)$.

\subsection{Non-local Constraint Checking}
\label{sxn:ca_nlcc}
We study the complexity of the NLCC routine for checking a single constraint $\cC_0 \in \cK_0$, presented in Alg.~\ref{alg:pc_2}. Note that for a cyclic constraint, a token must be initiated from each
vertex in the background graph that may participate in the substructure representing $\cC_0$, i.e., in Alg.~\ref{alg:pc_2}, each vertex in $\cG^*$, that match at least one vertex in $\cC_0$, initiates a token.
\\

\textit{\textbf{Space Complexity.}} 
The NLCC routine requires two additional algorithm states: 
\textit{(i)} $\gamma$ - the map of token source vertices (in $\cG^*$) for $\cC_0$, requires at most $O(|\cV^*|)$ storage. \textit{(ii)} $\tau(v_j)$ - the set of already forwarded tokens by a vertex $v_j$ used for work aggregation: if $\cC_0$ is edge-monocyclic and has unique vertex labels, per-vertex storage requirement for $\tau(v_j)$ is no more than $O(|\gamma|)$ or total $O(|\cV^*|\times|\gamma|)$ for $\cG^*$. For arbitrary templates, however, the cost is superpolynomial and proportional to the message complexity discussed later. Similarly, the worst case storage requirement for the visitor queue is also superpolynomial (and directly related to the generated message traffic).
\\

\textit{\textbf{Time Complexity.}} 
In NLCC, each constraint $\cC_0\in\cK_0$ is verified by passing around tokens. Each active vertex in $\cV^*\in\cG^*$ that could be a template match for the first vertex in $\cC_0$, issues a token - identified by an entry in $\gamma$ where $|\gamma|\leq|\cV^*|$. In the distributed message passing setting, token passing happens in a breadth-first search manner (on shared memory, a more work-efficient depth-first search like implementation is possible). The effort related to token propagation is bounded by $|\gamma|$ - the number of tokens, average degree connectivity, and the depth of the propagation (i.e., the size of the constraint $|\cC_0|$). For an arbitrary constraint $\cC_0$, the cost is exponential: assume $r$ indicates a step in the walk represented by $\cC_0$; at $r=1$, in the worst case, a token is received by at most $(|\cV^*|-1)$ vertices, and at $r=2$, each of these vertices forward the same token to at most $(|\cV^*|-2)$ vertices. To propagate $|\gamma|$ token, this results in visiting $|\gamma|\times(|\cV^*|-1)\times(|\cV^*|-2) \times \dots \times (|\cV^*|-r-1)$ vertices, where $r=|\cC_0|)$. Since $|\gamma|\leq|\cV^*|$, we can write the sequential cost of verifying constraint $\cC_0$ is $O(|\cV^*|^{|\cC_0|})$. 
\\

\textit{\textbf{Message Complexity.}} 
As discussed above, in NLCC, each vertex visitation by (a copy of) a token results in one message. Therefore, the message complexity of checking a non-local constraint $\cC_0$ is $O(|\cV^*|^{|\cC_0|})$. Heuristics like work aggregation, however, prevents a vertex from forwarding duplicate copies of a token, which reduces the time and message propagation effort in practice. 


\subsection{Motivating the Expected Gains from the Complexity Perspective}
The previous section presents the time, space, and message complexity of the local and non-local constraint checking algorithms. Here, we attempt to give an intuition for the expected performance gains compared to the traditional 
direct enumeration approach ~\cite{Ullmann:1976:ASI:321921.321925}.
Direct enumeration has $O(|\cV|^{|\cG_0|})$ complexity in the general case ~\cite{Ullmann:1976:ASI:321921.321925}. In our approach, the non-local constraint checking routines are the high-complexity routines: $O(|\cV^*|^{|\cC_0|})$. These routines operate on the current solution subgraph graph $\cG^*(\cV^*,\cE^*)$ after it has already been pruned by local constraint , and is generally expected to be significantly smaller than the original background graph, i.e., $|\cV^*|\leq|\cV|$ (we explore this in ~\cite{Reza:2018:SC:PM}; note that we eliminate both vertices and edges). Also, $|\cC_0|\leq|\cG_0|$ and, as we check constraints in the increasing order of their length, constraints (substructures of the search template) that require a longer walk, operate on the smaller pruned graph available in the later stages of processing. Finally, compared to direct enumeration, 
our constraint checking based approach typically generates smaller algorithm state - thus limiting combinatorial explosion; and, at the same time, the work aggregation heuristic prevents a vertex from forwarding duplicate copies of a token, which reduces the generated network traffic (see \S\ref{snx:optimization_evaluation}). In the same vein, in our approach, match enumeration is performed on the pruned solution  subgraph, hence, the complexity is $O(|\cV^*|^{|\cG_0|})$.


\section{Evaluation}
\label{sxn:evl}

\noindent
This section is structured as follows: To demonstrate the ability of our system to process massive graphs on large deployments, we present \emph{strong scaling} experiments on the largest real-world graph publicly available (\S\ref{sxn:evl_strong_scaling}). 
We evaluate the effectiveness of key design decisions, optimizations, and load balancing techniques our system incorporates (\S\ref{snx:optimization_evaluation} and \S\ref{sxn:load_balancing_evaluation}). We demonstrate the versatility of our constraint checking approach and use it as a stepping stone to efficiently support additional usage scenarios, 
namely, \emph{interactive incremental search} and \emph{exploratory search}
(\S\ref{sxn:evl_usage_scenarios_8}).  
We compare our solution with three state-of-the-art exact pattern matching systems, Arabesque~\cite{Teixeira:2015:ASD:2815400.2815410}, QFrag~\cite{Serafini:2017:QDG:3127479.3131625} and TriAD~\cite{Gurajada:2014:TDS:2588555.2610511} (\S\ref{sxn:compare_others}). Furthermore, we study how search template characteristics impact search performance and (\S\ref{sxn:evl_template_sensitivity_analysis}) and 
demonstrate application to graphs with various vertex degree distributions (\S\ref{sxn:evl_graph_sensitivity_analysis}).

\indent Our previous work~\cite{Reza:2018:SC:PM}, includes additional experimental results: \emph{weak scaling} experiments on massive synthetic \emph{R-MAT} graphs with up to $\sim$4.4 trillion edges, and using  up to 1,024 compute nodes (36,864 cores) (\cite{Reza:2018:SC:PM}, \S5A); demonstrates the ability to support \emph{full match enumeration}, starting from the pruned solution subgraph (\cite{Reza:2018:SC:PM}, \S5A) on these massive datasets; evaluation of various design decisions (\cite{Reza:2018:SC:PM}, \S5F); shows support for realistic data analytics scenarios using two real-world graphs, \emph{Reddit} and \emph{IMDb} (\cite{Reza:2018:SC:PM}, \S5D); and an exploration of time-to-solution vs. precision guarantees trade-offs (\cite{Reza:2018:SC:PM}, \S5E). Finally, \cite{Nicolas:2018:IA3:PM:8638389} explores advanced heuristics for constraint selection and ordering; and ~\cite{Reza:2017:IEEECluster:PM.8048872} focuses on a restricted set of search templates, acyclic or edge-monocyclic without duplicate labels, that can be supported extremely efficiently.

\subsection{Testbed}
The testbed is the 2.6 petaflop Quartz cluster at the Lawrence Livermore National Laboratory, comprised of 2,634 nodes and the Intel Omni-Path interconnect. Each node has two 18-core Intel Xeon E5-2695v4 @2.10GHz processors and 128GB of main memory~\cite{quartz.001}. We run one MPI process per core (i.e., 36 processes per node).

\subsection{Datasets} 
Table~\ref{table:evl_datasets} summarizes the main characteristics of the datasets used in this work.  We briefly explain below how the background graphs and their labels are created. Additional details can be found in~\cite{Reza:2018:SC:PM,Nicolas:2018:IA3:PM:8638389}. For all graphs, we created undirected versions - two directed edges are used to represent each undirected edge.  

\begin{table}[ht!]
{
\footnotesize
\renewcommand{\arraystretch}{1.2}
\captionsetup{font=footnotesize}
\caption{Properties of the datasets used for evaluation: number of vertices and edges, maximum, average and standard deviation of vertex degree, and the graph dataset size, in the compact CSR-like representation used, which includes the vertex metadata.
}
\label{table:evl_datasets}
\begin{center}
\textcolor{black}{
\begin{tabular}{lcrrrrrr}
\hline
 & {Type} & {$|\cV|$} & {$2|\cE|$} & {$d_{max}$} & {$d_{avg}$} & {$d_{stdev}$} & {Size} \\ 
\hline
Web Data Commons \cite{wdc.2012} 
& Real
& 3.5B & 257B & 95M & 72.3 & 3.6K & 2.7TB \\
\hline
Reddit \cite{reddit.public.data.2017} & Real
& 3.9B &  14B & 19M & 3.7 & 483.3 & 460GB \\
\hline
Internet Movie Database \cite{imdb.public.data.2016} & Real
& 5M &  29M & 552K & 5.8 & 342.6 & 581MB \\
\hline
CiteSeer~\cite{Teixeira:2015:ASD:2815400.2815410} & Real
& 3.3K &  9.4K & 99 & 3.6 & 3.4 & 741KB\\
\hline
Mico~\cite{Teixeira:2015:ASD:2815400.2815410} & Real
& 100K & 2.2M & 1.4K & 22 & 37.1 & 36MB\\
\hline
Patent \cite{Serafini:2017:QDG:3127479.3131625} & Real
& 2.7M & 28M & 789 & 10.2 & 10.8 & 480MB\\
\hline
YouTube \cite{Serafini:2017:QDG:3127479.3131625} & Real
& 4.6M & 88M & 2.5K & 19.2 & 21.7 & 1.4GB \\
\hline
LiveJournal~\cite{Backstrom:2006:GFL:1150402.1150412} & Real
&  4.8M & 69M & 20K & 17 & 36 & 1.2GB\\
\hline
Twitter~\cite{Kwak:2010:TSN:1772690.1772751} & Real
&  41.7M & 2.9B & 3M & 47.7 & 2.1K & 47GB\\ 
\hline
UK Web~\cite{Boldi:2004:WFI:988672.988752,Boldi:2011:LLP:1963405.1963488} & Real
&  105.9M & 7.5B & 975K & 70.6 & 718 & 119GB\\ 
\hline
Road USA~\cite{Rossi.2015.AAAI.NDR} & Real
&  23.9M & 58M & 9 & 2.4 & 0.9 & 1.4GB\\
\hline
R-MAT up to Scale 37 \cite{Chakrabarti04r-mat:a} 
& Synthetic
& 137B & 4.4T & 612M & 32 & 4.9K & 45TB \\ 
\hline
\end{tabular}
}
\end{center}
}
\end{table}

The \textit{Web Data Commons (WDC)} graph is a webgraph whose vertices are webpages and edges are hyperlinks. To create vertex labels, we extract the top-level domain names from the webpage URLs, e.g., \emph{.org} or \emph{.edu}. If the URL contains a common second-level domain name, it is chosen over the top-level domain name. For example, from \emph{ox.ac.uk}, we select \emph{.ac} as the vertex label. A total of 2,903 unique labels are distributed among the 3.5B vertices in the background graph. 

We curated the \textit{Reddit (RDT)} social media graph from an open archive~\cite{reddit.public.data.2017} of billions of public posts and comments from Reddit.com. 
Reddit allows its users to rate (upvote or downvote) others' posts and comments. The graph has four types of vertices: \emph{Author, Post, Comment} and \emph{Subreddit} (a category for posts). For \emph{Post} and \emph{Comment} type vertices there are three possible labels: \emph{Positive, Negative,} and \emph{Neutral} (indicating the overall balance of positive and negative votes) or \emph{No} rating. An edge is possible between an \emph{Author} and a \emph{Post}, an \emph{Author} and a \emph{Comment}, a \emph{Subreddit} and a \emph{Post}, a \emph{Post} and a \emph{Comment} (to that \emph{Post}), and between two \emph{Comments} that have a parent-child relation.  


We use the smaller \textit{Patent} and \textit{YouTube} graphs for comparison with existing exact pattern matching systems, QFrag~\cite{Serafini:2017:QDG:3127479.3131625} and TriAD~\cite{Gurajada:2014:TDS:2588555.2610511}.
The \textit{Patent} graph has 37 unique vertex labels, while the \textit{YouTube} graph has 108 unique vertex labels. We use \textit{CiteSeer}, \textit{Mico}, \textit{Patent}, \textit{YouTube} and \textit{LiveJournal} unlabeled, real-world graphs for performance comparison with Arabesque~\cite{Teixeira:2015:ASD:2815400.2815410}. 
Additionally, we use  two large (billions of edges) real-world, scale-free graphs, \textit{Twitter} and \textit{UK Web}, used in the past by many for studying various graph analysis problems; and a large diameter, real-world, road network graph, \textit{Road USA}.

The synthetic \textit{Recursive MATrix (R-MAT)} graphs exhibit approximate power law degree distribution~\cite{Chakrabarti04r-mat:a}. These graphs were created following the Graph 500 \cite{graph500.2016} standards:  $2^{Scale}$  vertices and a directed edge factor of 16. For example, a Scale 30 graph has $|\cV|=2^{30}$ and $|\cE| \approx 32\times2^{30}$ (as we create an undirected version). Since we use the \emph{R-MAT} graphs for weak scaling experiments, we aim to generate labels such that the graph structure changes little as the graph scales. To this end, we leverage vertex degree information to create vertex labels, computed using the formula, $\ell(v_i) = \lceil \log_2(d(v_i)+1)\rceil$. This, for instance for the Scale 37 graph, results in 30 unique vertex labels.
\\

\textit{\textbf{Notes on Data Storage and Loading.}} Our testbed is served by a distributed storage platform running the Lustre parallel file system~\cite{Lustre.001}. To accelerate graph loading, HavoqGT can preprocess the adjacency lists to take advantage of the existing parallel file system: it splits each input dataset in the same number of parts/files as the MPI processes used in the respective experiment. HavoqGT's graph partitioning process also attempts to create balanced partitions by assigning an equal share of edges to each partition and, where necessary, splits the edge set of a high-degree vertex over multiple partitions. This however, can be a costly process for massive graphs: for example, for the \emph{WDC} graph, graph partitioning for 128 nodes (4,608 partitions) takes about six hours. This distributed graph can then be loaded from the parallel file system in under two minutes. The vertex metadata, is stored (split in multiple parts/files) independently of the graph topology and can be loaded from the distributed file system relatively fast without preprocessing: for example, for the \emph{WDC} graph, in about 30 seconds.

\subsection{Search Templates and Experiment Design} 
\begin{figure}[!t]
\centering
\includegraphics[width=3.7in]{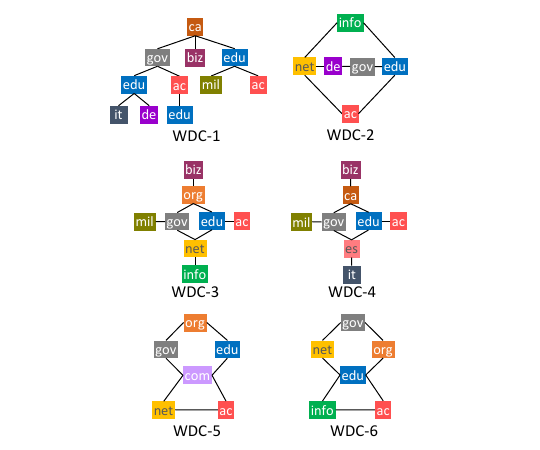} 
\setlength{\belowcaptionskip}{-12pt}
\captionsetup{font=footnotesize}
\caption{\textcolor{black}{\textit{WDC} patterns using top/second-level domain names as labels. The labels selected are among the most frequent, covering $\sim$81\% 
of the vertices in the \textit{WDC} graph: unsurprisingly, \emph{com} is the most frequent - covering over two billion vertices, \emph{org} covers $\sim$220M vertices, the 2\textsuperscript{nd} most frequent after \emph{com} and \emph{mil} is the least frequent among these labels, covering $\sim$153K vertices.}}
\label{fig:wdc_0011}
\end{figure}

To stress our system, we use templates based on patterns naturally occurring, and relatively frequent, in the background graphs. The \textit{WDC} (Fig.~\ref{fig:wdc_0011}), \textit{Twitter}, \textit{UK Web}, \textit{Patent}, \textit{YouTube} (Fig.~\ref{fig:qfrag_patterns_0011}) and \textit{R-MAT} patterns include vertex labels that are among the most frequent in the respective graphs. The \textit{Reddit} and \textit{IMDb} patterns 
include most of the vertex labels in these two graphs~\cite{Reza:2018:SC:PM}. We chose templates to exercise different constraint checking scenarios: the search templates have multiple vertices with the same label and non-edge-monocyclic properties (they require relatively expensive non-local constraint checking). 

All runtime numbers provided are averages over 10 runs. 
Unless mentioned explicitly, the performance metric is the time to produce the solution subgraph for a single template. 

\begin{figure}[!t]
\centering
\includegraphics[width=5.0in]{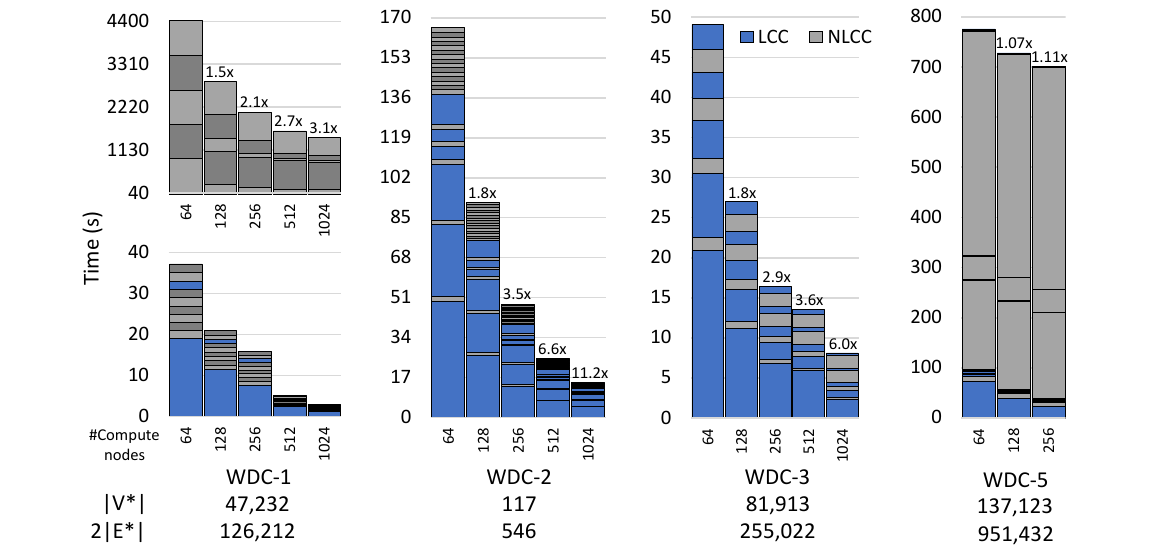} 
\captionsetup{font=footnotesize}
\caption{Runtime for strong scaling experiments, broken down into individual (LCC and NLCC) phases, 
for four of the patterns in Fig.~\ref{fig:wdc_0011}. (For better visibility, for WDC-1, 
runtime for different iterations are split into two scales on the Y-axis.) 
The last two rows are the number of vertices and edges in the pruned solution subgraph, respectively. Speedup over the 64 node configuration is also shown on top of each stacked bar plot. 
(Partially reused from~\cite{Reza:2018:SC:PM} with additional results.)}
\label{fig:strong_scaling_0011}
\end{figure}

\subsection{Strong Scaling Experiments}
\label{sxn:evl_strong_scaling}
\noindent
The strong scaling experiments evaluate the performance of pruning (i.e., we verify all the constraints required to guarantee zero false positives). The smallest experiment uses 64 nodes, as this is the lowest number of nodes that can load the graph topology and vertex metadata in memory.  Fig.~\ref{fig:strong_scaling_0011} shows runtimes for strong scaling experiments when using the real-world \textit{WDC} graph on up to 1,024 nodes (36,864 cores). Intuitively, pattern matching on the \textit{WDC} graph is harder than on the \textit{R-MAT} graph as the \textit{WDC} graph is denser, has a highly skewed degree distribution, and the high-frequency labels used also belong to vertices with high neighbor degree.

We use the patterns presented in Fig.~\ref{fig:wdc_0011}. WDC-1 is acyclic, yet has multiple vertices with the same label and thus requires non-local constraint checking (PC and TDS). For better visibility, the plot splits checking initial LCC and NLCC-path constraints (bottom left) from NLCC-TDS constraints (top left). We notice near perfect scaling for the LCC phases, however, some of the NLCC phases do not show linear scaling (explained in \S\ref{sxn:load_balancing_evaluation}). 

WDC-2 is an example of a pattern with multiple cycles sharing edges, and relies on CC and TDS constraint checking. WDC-2 shows near-linear scaling with $\sim$1/3 of the total time spent in the first LCC phase and little time spent in the NLCC phases. WDC-3 is a monocyclic template and, when edge elimination is used (bottom right), shows steady scaling for both LCC and NLCC phases. 

The WDC-5 pattern includes the top three most frequent labels, namely, \emph{com}, \emph{org} and \emph{net}, and covers $\sim$72\% vertices in the \emph{WDC} graph. Similar to WDC-1, a majority of the time is spent verifying the non-local constraints. The NLCC phases do not scale well with increasing node count for two interrelated reasons: first, vertices participating in matches have high neighbor degree, and second, and more importantly, heavily skewed template match distribution among the graph partitions, (further explored in \S\ref{sxn:load_balancing_evaluation}).

\subsection{Impact of Major Design Decisions and Optimizations}
\label{snx:optimization_evaluation}
Here, we present the impact of two major design decisions and optimizations: \textit{(i)} Edge Elimination, and \textit{(ii)} Work Aggregation. 
(In~\cite{Reza:2018:SC:PM} and~\cite{Nicolas:2018:IA3:PM:8638389}, we have studied the impact of additional design features on search performance.)  
\\

\begin{figure}[!t]
\centering
\subfloat[]{\includegraphics[width=1.8in]{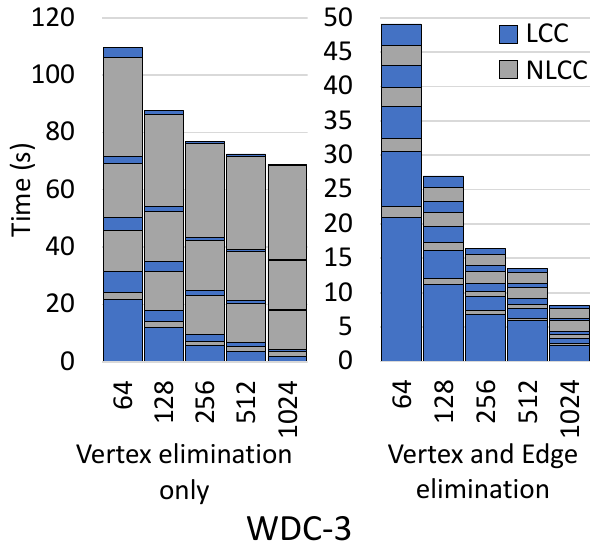}}
\hspace{12pt}
\hspace{12pt}
\subfloat[]{\includegraphics[width=1.8in]{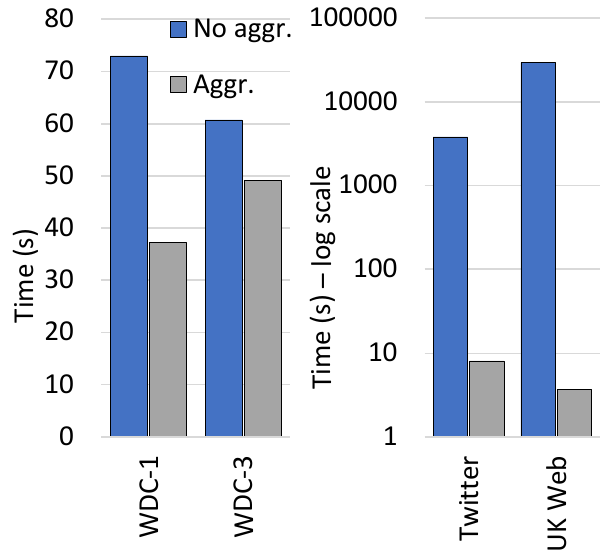}} 
\hspace{12pt}
\captionsetup{font=footnotesize}
\caption{\textcolor{black}{(a) Performance and scalability comparison between the vertex elimination only (left), and combined vertex and edge elimination (right), for the WDC-3 pattern. 
(b) Impact of work aggregation on runtime using three real-world graphs: WDC-1 and WDC-3 patterns (for the sake of readability, only a subset of non-local constraints are considered for WDC-1), and the Q8 pattern (Fig.~\ref{fig:qfrag_patterns_0011}) using the \emph{Twitter} and \emph{UK Web} graphs.  
(Partially reused from~\cite{Reza:2018:SC:PM} with additional results.) 
}}
\label{fig:perf_comp_0011}
\end{figure}

\textit{\textbf{Edge Elimination.}} Fig.~\ref{fig:perf_comp_0011}(a) highlights the important scalability and performance impact of edge elimination: without it, the NLCC phases take almost one order of magnitude longer and the entire pruning takes 2--9$\times$ longer. Without edge elimination, the WDC-3 pattern results in 3,180,678 edges selected (it includes false positives). Edge elimination identifies the true positive matches and reduces the number of active edges to 255,022. In other words, the solution subgraph is 12.5$\times$ sparser which in turn improves overall message efficiency of the system. We note that this one order of magnitude reduction enables match enumeration and advanced analytics on the solution subgraph.
\\

\textit{\textbf{Work Aggregation.}} Fig.~\ref{fig:perf_comp_0011}(b) shows the performance gains enabled by the work aggregation strategy employed by the distributed non-local constraint checking routine (presented in \S\ref{snx:di} and Alg.~\ref{alg:pc_1}). We study the impact of work aggregation for three large real-world graphs: \emph{WDC}, \emph{Twitter} and \emph{UK Web}. The magnitude of the gain 
is data dependent and more pronounced when the pattern is abundant, e.g., 50\% improvement for WDC-1 which has 600M+ matches in the background graph. The experiments using the \emph{Twitter} and \emph{UK Web} graphs further highlight the advantage of work aggregation: we compare the runtime of a single non-local constraint (a TDS constraint involving all the vertices and edges in the template) for the search pattern Q8 (Fig.~\ref{fig:qfrag_patterns_0011}). (The experiment details are available in \S\ref{sxn:evl_template_sensitivity_analysis}.) For the \emph{Twitter} and \emph{UK Web} graphs, the gain in runtime are two and three orders of magnitude, respectively (Fig.~\ref{fig:perf_comp_0011}(b), right chart). Unlike full match enumeration, NLCC does not need to identify all possible walks for each token; the goal is to identify the existence of any such walk (a complete path) in the background graph - sufficient to save the vertex that initiated the token from elimination. The significant improvement in runntime is due to reduction in number of complete paths traversed by all the tokens created; the number of messages communicated in non-local constraint checking is proportional to the number of paths traversed. For the \emph{UK Web} graph, for 24,000 unique tokens, without work aggregation, 45 billion unique paths are discovered. Our work aggregation technique reduces the number of complete paths traversed to 71 million, 
a four orders of magnitude reduction; hence, the three orders of magnitude gain in runtime (Fig.~\ref{fig:perf_comp_0011}(b), right chart). Similarly, for the Twitter graph, the reduction in the number of complete paths traversed is three orders of magnitude. (In Fig.~\ref{fig:perf_comp_0011}(b), all experiments were run on 64 compute nodes.)

\subsection{Load Balancing}
\label{sxn:load_balancing_evaluation}
For our pattern matching solution, load imbalance can be caused by two artifacts: First, over the course of execution our solution causes the workload to mutate, as it prunes away vertices and edges. Second, the distribution of matches in the background graph may be nonuniform: matches may reside on a small, potentially concentrated, portion of the graph. This section, first presents a detailed characterization of these artifacts, then it discusses and evaluates two load balancing strategies.
\\

\textit{\textbf{Does Load Imbalance Occur?}} 
Indeed, load imbalance does occur. For instance, for the relatively rare WDC-2 (Fig.~\ref{fig:wdc_0011}) pattern, when using 64 nodes, for example, the vertices and edges that participate in the final selection are distributed over as few as 111 partitions out of the 2,304 (64 nodes $\times$ 36 MPI processes per node). The distribution is concentrated - more than half of the matching edges reside on only 20 partitions. For the more frequent WDC-1 pattern, 50\% of the matching edges are on less than 5\% of the partitions on a 64 node deployment, and less than 3\% of the partitions on a 128 node deployment.

We observe further nonuniformity in the match distribution at the vertex granularity; the number of matches a vertex participates in, can significantly vary across the matching vertex set $\cV^*$. As an example, let's consider the WDC-2 pattern (whose matches are shown in~\cite{Reza:2018:SC:PM}, Fig. 10; they form six connected components). The largest connected component contains 2,262 matches (bottom row, center). In this connected component, there is a single \emph{gov} vertex, which participates in 2,262 matches (out of a total of 2,444 matches). This artifact is more pronounced in the case of the WDC-1 and WDC-2 patterns. For WDC-1, 99\% of the matching vertices are part of a single connected component. There are multiple vertices that belong to over three million matches. The numbers are more striking for the frequent WDC-3 pattern - a single vertex participates in over 34 million matches.

This irregularity has crucial performance implications, in particular, it hinders the scalability of the routines that rely on \emph{multi-hop graph walks}, such as non-local constraint checking and full match enumeration. When the matches are concentrated on a few compute nodes and only a few vertices participate in a large number of matches, the partitions these vertices reside on send/receive a larger portion of the message traffic. In this case, increasing the number of processors does not help as, in our current infrastructure, processing at the vertex granularity can not be `scaled out' efficiently. Furthermore, since each partition processes the local message queue sequentially, message traffic targeting popular vertices can overwhelm the respective partitions. Consequently, these bottlenecked partitions become the key performance limiter. This reasoning explains why some of the non-local constraint checking phases do not scale well (e.g.,  Fig.~\ref{fig:strong_scaling_0011}). 
\\

\textit{\textbf{Strategies to Address Load Imbalance Issues.}} We explore two strategies to address load-balancing issues: \textit{(i)} reshuffling the load, and \textit{(ii)} load consolidation, i.e., reloading the shuffled load on fewer nodes to also optimize for locality and reduce generated network traffic.
\\

\textit{Load Reshuffling.} We employ a \emph{pseudo-dynamic}, load balancing strategy. First, we checkpoint the current state of execution: the pruned graph, i.e., the set of active vertices and edges and the per-vertex state indicating template matches, $\omega(v_j)$ (Alg.~\ref{alg:vertex_state_1}). Next, using HavoqGT's graph partitioning module, we reshuffle the vertex-to-processor assignment to evenly distribute vertices (with $\omega(v_j)$ remained intact) and edges across processing cores. Processing is then resumed on the rebalanced workload. Depending on the size the the pruned graph, it is possible to resume processing on a smaller deployment (discussed in the next section). Over the course of the execution, checkpointing and rebalancing can be repeated as needed (the identification of the optimal trigger point to perform load balancing, however, requires further investigation).

\begin{figure}[!t]
\centering
\includegraphics[width=5.5in]{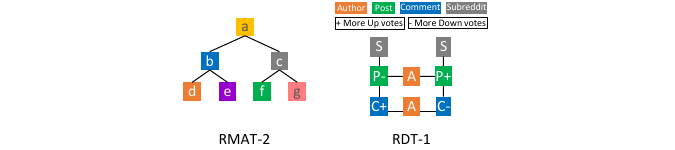}
\captionsetup{font=footnotesize}
\caption{\textcolor{black}{RMAT-2 is the template used for \emph{R-MAT} experiments (left); it includes the most frequent vertex labels in the background
graph (
as in Fig.~\ref{fig:qfrag_patterns_0011}). The RDT-1 (\emph{Reddit}) pattern is used for the load balancing experiments in \S\ref{sxn:load_balancing_evaluation} (right, reused from~\cite{Reza:2018:SC:PM}).}}
\label{fig:rmat_2_0014}
\end{figure}

\begin{figure}[!t]
\centering
\subfloat[Load balancing]{\includegraphics[width=1.5in]{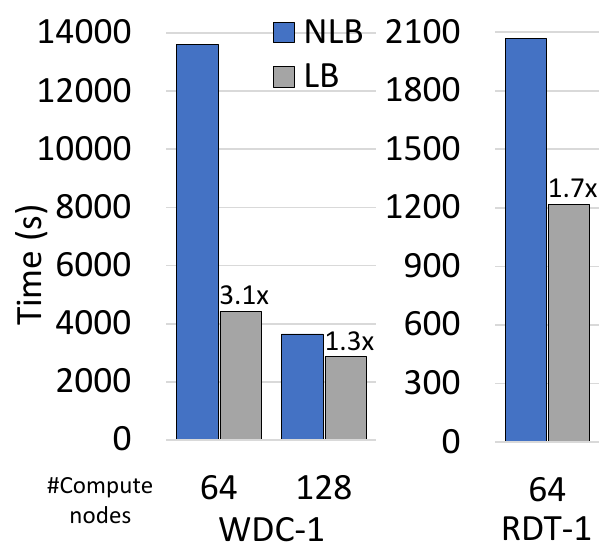}}
\hspace{12pt}
\subfloat[Time vs. CPU Hour]{\includegraphics[width=1.5in]{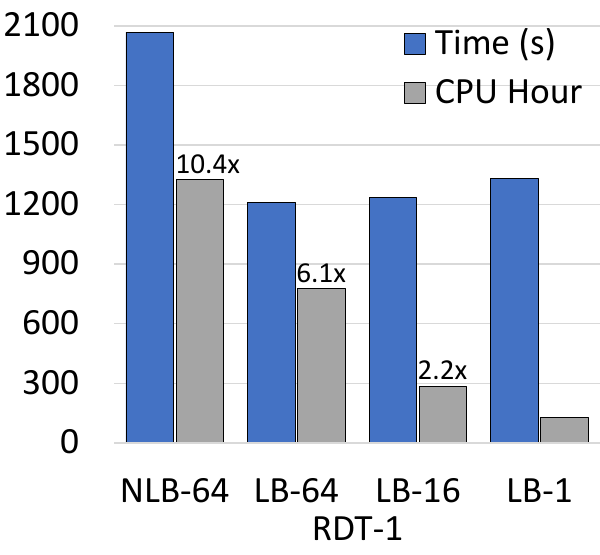}}
\captionsetup{font=footnotesize}
\caption{(a) Impact of load balancing on runtime using differnt number of compute nodes for the the WDC-1 and RDT-1 patterns. 
Speedup achieved by LB (load balancing) over NLB (without load balancing) is also shown on the top of each bar. 
(b) Performance of RDT-1 for four scenarios: (left) without load balancing on 64 nodes (NLB-64), (center-left) with load balancing 
on the same number of nodes (LB-64), (center-right) 
relaunching on a 16 node deployment after load balancing (LB-16) and (right) relaunching on a single node (36 processes) after load balancing (LB-1). 
Time-to-solution and CPU Hours consumed (normalized to the LB-1 experiment) are numerically presented on the top of the respective bars.
}
\label{fig:load_balancing_0011}
\end{figure}

To examine the impact of this technique, we analyze the runs for WDC-1 (Fig.~\ref{fig:wdc_0011}) and RDT-1 (Fig.~\ref{fig:rmat_2_0014}) patterns, as real-world workloads are more likely to lead to imbalance.  Fig.~\ref{fig:load_balancing_0011}(a) compares performance with and without load balancing. For these experiments, we perform rebalancing only once. For WDC-1, before verifying the TDS constraints, and for RDT-1, when the pruned graph is four orders of magnitude smaller. The extent of load imbalance is more severe for WDC-1 on the smaller 64 node deployment compared to using 128 nodes - workload rebalancing improves time-to-solution by 3.1$\times$ and 1.3$\times$, on 64 and 128 nodes, respectively. In the case of RDT-1, load balancing improves time-to-solution by 1.7$\times$. 
Given the pruned graphs are much smaller than the original graph; often the time spent in checkpointing, rebalancing, and relaunching the computation is negligible compared to the gain in time-to-solution. 
\\

\textit{Smaller Deployment.} One may argue that when the current solution subgraph $\cG^*$ is sufficiently small, it is more efficient to create load balanced partitions targeting a smaller deployment. Two different aspects of `efficiency' concerns support this approach: First, moving to a smaller deployment reduces power usage and may yield better normalized performance with respect to energy consumption. Second, for the scenario where the matches are highly concentrated on a limited number of nodes/partitions (which hinders the scalability of the non-local constraint checking phase), a smaller deployment offers locality (through reduced number of generated network traffic).

We setup a simple case study using the \textit{Reddit} dataset and the RDT-1 pattern. We resume processing on the rebalanced workload on a smaller deployment - from the original 64 node deployment, we switch to a 4$\times$ smaller deployment comprised of 16 nodes. In a second use case, we resume processing on the rebalanced workload a on a single node (running 36 processes). Fig.~\ref{fig:load_balancing_0011}(b) compares four scenarios: \textit{(i)} without load balancing (NLB-64), \textit{(ii)} with load balancing (LB-64), \textit{(iii)} with load balancing and relaunching on a smaller 16 node deployment (LB-16), and \textit{(iv)} relaunching on a single node after load balancing (LB-1). In addition to time-to-solution, we also compare CPU Hours consumed by each of the four cases. (A platform's net energy consumption can be roughly approximated by the total CPU Hours expended.) Fig.~\ref{fig:load_balancing_0011}(b) shows that, with respect to time-to-solution, LB-64 has marginal advantage over LB-16 and LB-1. However, LB-1 holds significant advantage in terms CPU Hour consumption: it is 6.1$\times$ more efficient than LB-64. The overhead for NLB-64 is 
10.4$\times$ compared to LB-1. These results support the argument that the load balanced partitions targeting a smaller deployment yields better normalized performance with respect to CPU Hour and energy consumption. 

\subsection{Advanced Usage Scenarios}
\label{sxn:evl_usage_scenarios_8}
This section highlights that our approach, based on constraint checking, 
can be extended to support a number of advanced pattern matching scenarios.
\\

\begin{figure}[!t]
\centering
\includegraphics[width=4.0in]{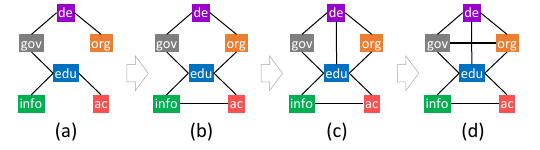} 
\captionsetup{font=footnotesize}
\caption{\textcolor{black}{An example showing the queries incrementally searched in the \textit{WDC} graph. The user begins with the left most pattern, (a), and incrementally revises the query by adding edges; gradually moves from left to right.  
}
}
\label{fig:wdc_0091}
\end{figure}

\begin{figure}[!t]
\centering
\includegraphics[width=2.0in]{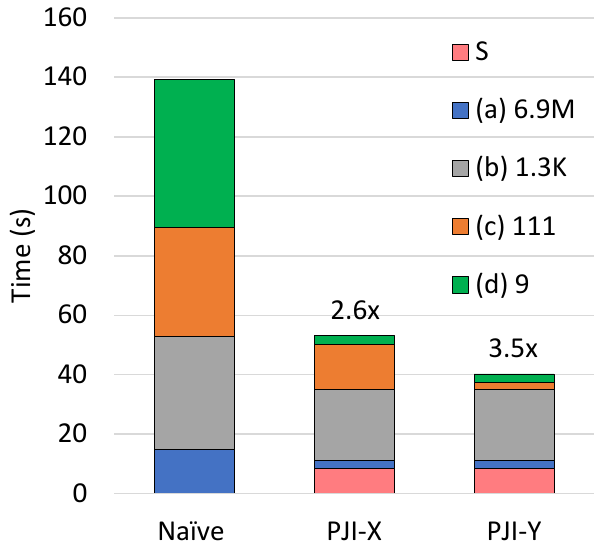} 
\captionsetup{font=footnotesize}
\caption{Runtime comparison between the na\"ive and the optimized incremental search solution for the scenario in Fig.~\ref{fig:wdc_0091}.  
For each experiment, the stacked bar plot shows the time spent in each query. 
For the optimized incremental search solution (labeled PJI), we consider two setups: \textit{(i)} PJI-X - we build the candidate set, directly search the initial query in it, then each of the remaining searches is limited within the vertex set of the solution subgraph of the previous search; \textit{(ii)} PJI-Y - in addition to PJI-X, we also employ the work reuse technique. 
For PJI, speedup achieved over the na\"ive approach is shown on top of respective bar plots. The chart legend also shows the number of vertices that match respective queries. The fraction of the runtime labeled `S' is the overhead for building the candidate set. (Partially reused from~\cite{Reza.2020.GRADES.10.1145/3398682.3399166}.)
}
\label{fig:wdc_incremental_search_0011}
\end{figure}

\textit{\textbf{Interactive Incremental Search.}} 
In this scenario, the user starts with an under-constrained search template (possibly returning too many matches), and the system is setup for interactive use: the user can add/delete edges, observe the changes in the solution subgraph (or statistic over it), and continue to interact with the system. The only restriction we place on the user is that (s)he can remove only edges (not vertices) and has to maintain the search template is connected. 

We take advantage of two observations: \textit{(i)} For template revision through edge addition, adding an edge is similar to adding a constraint, and the search for the revised template can be limited within the vertex set of the current solution subgraph presented to the user. 
\textit{(ii)} For template revision through edge deletion, we observe that, one can build a solution superset that is the union of all matches for all possible search templates that may be obtained from the initial template by removing just edges, using local constraints only, thus at a low cost. We use this restricted solution superset, 
which we refer to as the \textit{candidate set}, to initially prune the backgtound graph, and to reinitiate the solution subgraph when an edge (from the current search template) is deleted. Furthermore, for the same vertex in the background graph, non-local constraints can be verified only once and this information can be reused in later searches (i.e., for revised templates); eliminating a large amount of potentially redundant work. We call this technique \textit{work reuse}. (Design details are available in~\cite{Reza.2020.GRADES.10.1145/3398682.3399166}.)

For a preliminary evaluation of the effectiveness of these techniques we consider the search scenario presented in Fig.~\ref{fig:wdc_0091} and explained in detail in Fig.~\ref{fig:wdc_incremental_search_0011}. We demonstrate the advantage of the optimized technique over a nai\"ve approach that uses the exact matching solution to independently search the original query and each of its revisions. The experiment scenario, labelled PJI-X, highlights the advantage building and restricting the searches to the candidate set: the solution first computes the candidate set, and then each template revisions is searched starting from this set; this version of the optimized pipeline offers 2.6$\times$ speedup. The experiment scenario PJI-Y, in addition to computing the candidate set at the start of the experiment, also employs the work reuse technique, to eliminate redundant non-local constraint verification. This yields a further 3.5$\times$ gain in time-to-solution over the na\"ive approach. We run these experiments on 128 compute nodes  (4,608 cores).  
\\

\begin{figure}[!t]
\centering
\subfloat[]{\includegraphics[width=2.2in]{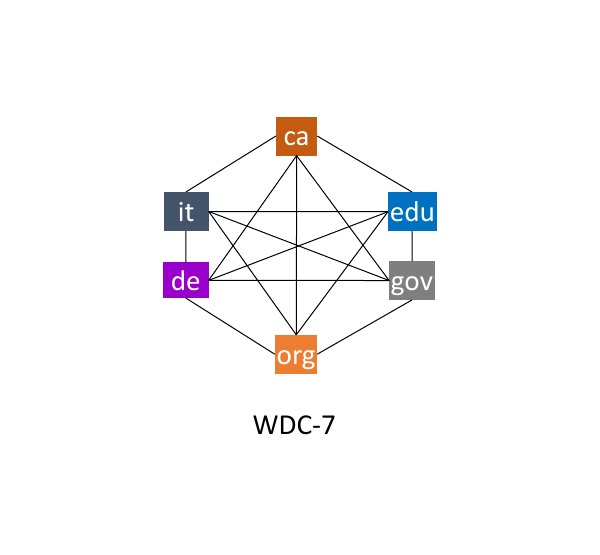}}
\subfloat[]{\includegraphics[width=2.2in]{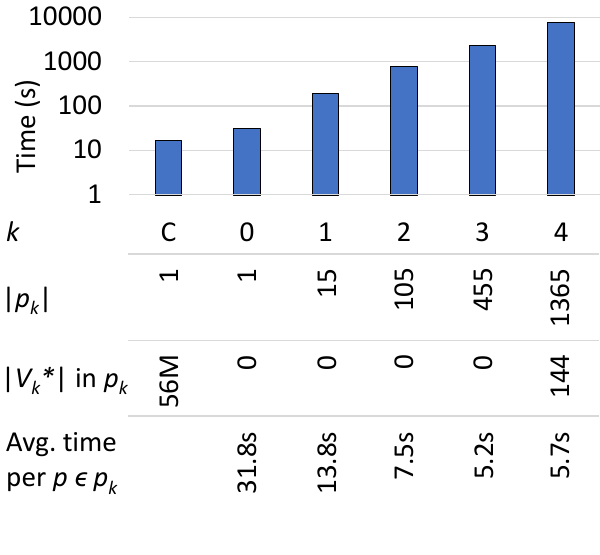}}
\captionsetup{font=footnotesize}
\caption{\textcolor{black}{(a) The \emph{WDC} pattern used for demonstration of exploratory search in \S\ref{sxn:evl_usage_scenarios_8}. (b) Runtime for WDC-7, grouped by multiple levels depending on the number of edges removed from the initial search template.  Note that no match is found until $k=4$ edges are removed. X-axis labels: (first row) $k$ is the number of edges removed.  (second row) $|p_k|$ is the number of distinct patterns that exist at level $k$; (third row) $|V^*_k|$ is the number of vertices that participate in any match of a pattern $p \in p_k$; (bottom row) average search time per pattern at each $k$. X-axis label `C' represents the initial candidate set generation. Note that the Y-axis is in log scale.}
}
\label{fig:apm_wdc_7_0014}
\end{figure}

\textit{\textbf{Exploratory Search.}} 
We present an exploratory search scenario where the user starts from an over-constrained search template and the system progressively relaxes the template by removing edges until matches are found. The search progresses as follows:  first, all variations of the initial search template with one edge removed are searched; then all variations with two edges removed are searches, and so on; until matches for at least one pattern are found.  The system returns a subgraph that is the union of all matches at the first level where matches are found. As in the case of interactive incremental search described earlier, here, the key enabler is to identify the candidate set and use this reduced set in the later iterations of the search, as well as reuse the result of non-local constraint checking. 

Fig.~\ref{fig:apm_wdc_7_0014}(b) shows the runtime (when using 128 nodes), broken down to each level, for such a search, in the \emph{WDC} graph, starting from an undirected 6-Clique (WDC-7 pattern in Fig.~\ref{fig:apm_wdc_7_0014}(a)). For this search template, the first matches are found only after four edges are removed and involves sifting through over 1,900 variations of the original search template - only 144 vertices participate in these matches. Note that reducing the search space to the candidate set and the high efficiency of the exact matching pipeline (on average, it takes less than six seconds to explore each variant of the template), enables this type of exhaustive search.


\subsection{Comparison with State-of-the-Art Systems}
\label{sxn:compare_others}
\noindent
We empirically compare our work with three state-of-the-art pattern matching systems QFrag~\cite{Serafini:2017:QDG:3127479.3131625}, TriAD~\cite{Gurajada:2014:TDS:2588555.2610511}, and Arabesque~\cite{Teixeira:2015:ASD:2815400.2815410}. QFrag is a generic pattern matching system; we use it for comparison using labeled patterns. Arabesque, although requires effort for writing pattern search algorithms, has demonstrated the ability to scale to much larger graphs; we use this system for comparison using unlabeled patterns. Since both QFrag and Arabesque are based on Apache Spark~\cite{Spark.001} and inherit its limitations; we also compare with an MPI-based solution, TriAD\footnote{Although TriAD is an RDF query processing engine and follows a distributed join based design, it has been shown to perform well for scale-free graphs~\cite{Serafini:2017:QDG:3127479.3131625}. Furthermore, TriAD is the only well performing MPI-based exact matching solution that is publicly available for evaluation.}. Similar to our system, all these three systems offer exact matching with 100\% precison and 100\% recall. For all experiments, we report time for a single query. We do not report time spent in graph loading and partitioning, and preprocessing (such as index creation in TriAD), as they are done once for each graph dataset, but we note that our system performs better or as well as the other systems. 

We run these experiments using real-world graph datasets, on a large shared memory platform: the machine is equipped with four Intel Xeon E7-4870v2 @2.30GHz CPU-sockets - a total of 60 CPU-cores and 120MB L3 memory, and 1.5TB main memory. For QFrag and Arabesque, we deploy HDFS on the local SAS disk array (in RAID-5).

\subsubsection{Comparison with QFrag}
\label{sxn:compare_qfrag}
Similar to our solution, QFrag targets exact pattern matching on distributed platforms, yet there are two main differences: QFrag assumes that the entire graph fits in the memory of each compute node and uses data replication to enable parallelism. More importantly, QFrag employs a sophisticated load balancing strategy to achieve scalability. QFrag is implemented on top of Apache Spark and Giraph~\cite{Giraph.001}. In QFrag, each replica runs an instance of the Turbo\textsubscript{ISO}~\cite{Han:2013:TIT:2463676.2465300} pattern enumeration algorithm (essentially an improvement of Ullmann's algorithm~\cite{Ullmann:1976:ASI:321921.321925}). Through evaluation, the authors demonstrated QFrag's performance advantages over two other distributed pattern matching systems: \textit{(i)} TriAD~\cite{Gurajada:2014:TDS:2588555.2610511} (which we confirm), and \textit{(ii)} GraphFrames~\cite{Dave:2016:GIA:2960414.2960416, GraphFrames.001}, a graph processing library for Apache Spark, also based on distributed join operations. 

\begin{figure}[!t]
\centering
\includegraphics[width=3.3in]{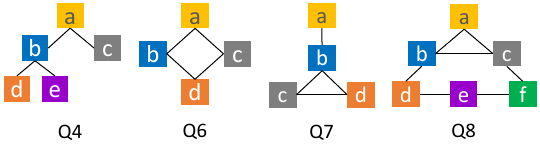} 
\captionsetup{font=footnotesize}
\caption{The patterns (reproduced  from~\cite{Serafini:2017:QDG:3127479.3131625}) used for comparison with QFrag and TriAD (results in Table~\ref{table:comparison_with_others}). The label of each vertex is mapped, in alphabetical order, to the most frequent label of the graph in decreasing order of frequency. Here, \emph{a} represents the most frequent label, \emph{b} is the second most frequent label, and so on.}
\label{fig:qfrag_patterns_0011}
\end{figure}

Given that we have demonstrated the scalability of our solution (Serafini et al. demonstrate equally good scalability properties for QFrag~\cite{Serafini:2017:QDG:3127479.3131625}, yet on much smaller graphs), we are interested to establish a comparison baseline at the single node scale. To this end, we run experiments on a modern shared memory machine with 60 CPU-cores, and use the two real-world graphs, \textit{Patent} and \textit{YouTube}, and four query patterns (Fig.~\ref{fig:qfrag_patterns_0011}) that were used for evaluation of QFrag~\cite{Serafini:2017:QDG:3127479.3131625}. We run QFrag with 60 threads and HavoqGT with 60 MPI processes. The results are summarized in Table~\ref{table:comparison_with_others}: QFrag runtimes for match enumeration (first pair of columns) are comparable with the results presented in~\cite{Serafini:2017:QDG:3127479.3131625}, so we have reasonable confidence that we replicate their experiments well. With respect to combined pruning and enumeration time, our system (second pair of columns, labeled PruneJuice MPI, presenting pruning and enumeration time separately) is consistently faster than QFrag on all the graphs, for all the queries. We note that our solution does not take advantage of shared memory of the machine at the implementation level (we use different processes, one MPI process per core), and has the system overhead of MPI-based communication between processes. Additionally, unlike QFrag, our system is not handicapped by the memory limit of a single machine as it supports graph partitioning and can process graphs larger than the main memory of a single node. 

\begin{table}[!t]
\footnotesize
\renewcommand{\arraystretch}{1.2}
\captionsetup{font=footnotesize}
\caption{\textcolor{black}{Performance comparison between QFrag, TriAD and PruneJuice using the patterns in Fig.~\ref{fig:qfrag_patterns_0011}. The table shows the runtime in seconds for full enumeration for QFrag and TriAD. 
For PruneJuice, we split time-to-solution into pruning time (top row) and enumeration time (bottom row). 
The best distributed runtime for a query, for each graph, is shown in bold font.
}}
\label{table:comparison_with_others}
\centering
\setlength\tabcolsep{2.5pt}
\textcolor{black}{
\begin{tabular}{c|c|c|c|c|c|c|c|c}
\hline
\multirow{2}{*}{} & \multicolumn{2}{c|}{QFrag} & 
\multicolumn{2}{c|}{TriAD} &
\multicolumn{2}{c|}{PruneJuice MPI} & \multicolumn{2}{c}{PruneJuice OpenMP}
\\
\cline{2-9}
& \multicolumn{1}{c|}{Patent} & \multicolumn{1}{c|}{YouTube} & 
\multicolumn{1}{c|}{Patent} & \multicolumn{1}{c|}{YouTube} &
\multicolumn{1}{c|}{Patent} & \multicolumn{1}{c|}{YouTube} & \multicolumn{1}{c|}{Patent} & \multicolumn{1}{c}{YouTube}
\\
\hline
\multirow{2}{*}{Q4} & 
\multirow{2}{*}{4.19} & 
\multirow{2}{*}{8.08} & 
\multirow{2}{*}{12.24} & 
\multirow{2}{*}{43.93} & 
\textbf{0.238} & \textbf{0.704} & 0.100 & 0.400
\\
\cline{6-9}
&&&&& \textbf{0.223} & \textbf{1.143} & 0.010 & 0.010
\\
\hline
\multirow{2}{*}{Q6} & 
\multirow{2}{*}{5.99} & 
\multirow{2}{*}{10.26} & 
\multirow{2}{*}{\textbf{0.89}} & 
\multirow{2}{*}{16.49} & 
0.874 & \textbf{2.340} & 0.070 & 1.730
\\
\cline{6-9}
&&&&& 0.065 & \textbf{0.301} & 0.005 & 0.010
\\
\hline
\multirow{2}{*}{Q7} & 
\multirow{2}{*}{6.36} & 
\multirow{2}{*}{11.89} & 
\multirow{2}{*}{1.08} & 
\multirow{2}{*}{11.16} & 
\textbf{0.596} & \textbf{1.613} & 0.130 & 0.820
\\
\cline{6-9}
&&&&& \textbf{0.039} & \textbf{0.180} & 0.005 & 0.010
\\
\hline
\multirow{2}{*}{Q8} & 
\multirow{2}{*}{10.05} & 
\multirow{2}{*}{14.48} & 
\multirow{2}{*}{\textbf{0.93}} & 
\multirow{2}{*}{29.09} & 
0.959 & \textbf{2.633} & 0.100 & 1.370
\\
\cline{6-9}
&&&&& 0.049 & \textbf{0.738} & 0.001 & 0.010
\\
\hline
\end{tabular}
}
\end{table}

To highlight the effectiveness of our technique and get some intuition on the magnitude of the MPI overheads in this context, we implemented our technique for shared memory (we use OpenMP for parallelization) and present runtimes (when using 60 threads) for the same set of experiments in Table~\ref{table:comparison_with_others} (the two rightmost columns, labeled PruneJuice OpenMP). We notice up to an order of magnitude improvement in performance compared to the distributed implementation running on a single node.

In summary, our distributed (PruneJuice MPI) solution works about 4--10$\times$ faster than QFrag, and, if excluding distributed system overheads and considering the pruning time for the shared memory solution (PruneJuice OpenMP) and conservatively reusing enumeration runtime for the distributed solution, it is about 6--100$\times$ faster than QFrag.


\subsubsection{Comparison with TriAD}
\label{sxn:compare_triad}
TriAD~\cite{Gurajada:2014:TDS:2588555.2610511} is distributed RDF~\cite{RDF.W3C.003} engine, implemented in MPI, and based on an asynchronous distributed join algorithm which uses \emph{partitioned locality} based indexing. The Resource Description Framework (RDF), is a metadata/typed graph model~\cite{RDF.W3C.003}, where information is stored as a linked \emph{Subject-Predicate-Object} triple. The \emph{Subject}, \emph{Predicate} and \emph{Object} are essentially designated types for graph vertices (forming a triple) and the links between them are edges in the graph. An RDF SPARQL~\cite{SPARQL.1.0.W3C.003} query disassembles a search template into a set of edges and the final results are constructed through \emph{multi-way join} operations~\cite{Gurajada:2014:TDS:2588555.2610511}. 

TriAD's design follows the classical master-worker 
architecture at indexing time, but allows for a direct, asynchronous communication among the worker nodes 
at query processing time. TriAD's index structure is optimized for processing hash joins. TriAD employs \emph{hash-based sharding} for data partitioning and partitioning information in encoded into the triples; which enables locality awareness and allows potentially large number of concurrent join operations by multiple worker nodes without the need for remote communication. Furthermore, in TriAD, the master node maintains a global index statistic (collected at local index creation time on 
worker nodes). This information is used by the query plan generator: query optimization is informed by a unified cost model for optimizing relational join operations.

We run the same experimnets for TriAD as we did earlier for comparison with QFrag, on the same graph datasets, queries, and large shared memory platform. 
The experiment results are summarzed in Table~\ref{table:comparison_with_others}, in the columns next to the QFrag results. For the smaller, less skewed, and sparser \emph{Patent} graph, except for search template Q4, TriAD's performance is on par with the distributed implementation of PruneJuice (and better than QFrag). In all other cases, particularly for the more skewed and dense \emph{YouTube} graph, TriAD performs much worse.  For Q4 and Q8, TriAD is $\sim$5$\times$ and $\sim$2$\times$ slower than QFrag, and $\sim$20$\times$ and $\sim$9$\times$ slower than distributed PruneJuice.

Although at the implementation level TriAD shares some similarities with PruneJuice (e.g., it leverages asynchronous processing); in contrast to QFrag and PruneJuice, TriAD follows a different design philosophy - distributed hash join operations. Whereas the solution approach QFrag uses is can be categorized as graph exploration~\cite{Gurajada:2014:TDS:2588555.2610511,Abdelaziz:2017:SEC:3151106.3151109}, an our solution ads graph pruning based on constraint checking to this. As expected, high-level design decisions are key drivers for performance: 
although QFrag operates within a managed runtime environment (i.e., JRE - the Java Runtime Environment) that is slower than the native MPI/C++ runtime, and relies on TCP for remote communication, which again, is slower than MPI communication primitives (typically optimized to harness shared memory IPC); QFrag's design enables sophisticated load balancing which is crucial for achieving good performance in presence of often highly skewed real-world graphs. Our design, in addition to harnessing asynchronous communication and embracing horizontal scalability, offers aggressive search space pruning while maintaining small algorithm states to prevent combinatorial explosion; thus, able to scale to large graphs as well as has been demonstrated to be performant for relatively small datasets. TriAD, although implemented in MPI, the join-based design suffers in presence of larger graphs and patterns with larger diameter.

In a recent study, Abdelaziz et al.~\cite{Abdelaziz:2017:SEC:3151106.3151109} pointed out a key scalability limitation of TriAD: following distributed join operations, to enable parallel processing, TriAD needs to re-shard intermediate results if the sharding column of the previous join is not the current join column. This cost can be significant for large intermediate results with multiple attributes. Also, their analysis in~\cite{Abdelaziz:2017:SEC:3151106.3151109} shows the significant memory overhead of indexing in TriAD (often larger than the actual graph topology). Also, we noticed that the overhead of index creation increases with the graph size: index creation time for the \emph{Patent} graph is about 2.5 minutes which goes up to about 7.7 minutes for the larger \emph{Youtube} graph.


\subsubsection{Comparison with Arabesque}
\label{sxn:compare_arabesque}
Arabesque is a exact matching framework offering precision and recall guarantees, implemented on top of Apache Spark and Giraph~\cite{Giraph.001}. Arabesque provides an API based on the Think Like an Embedding (TLE) paradigm, which enables a user to express graph mining algorithms tailored for each specific search pattern, and a BSP implementation of the search engine. Arabesque replicates the input graph on all worker nodes, hence, the largest graph scale it can support is limited by the size of the main memory of a single node. As Teixeira et al.~\cite{Teixeira:2015:ASD:2815400.2815410} showed Arabesque's superiority over other systems: G-Tries~\cite{Ribeiro:2014:GDS:2589412.2589422} and GRAMI~\cite{Elseidy:2014:GFS:2732286.2732289}, we indirectly compare with these two systems as well.

For the comparison, we use the problem of \emph{counting cliques} in an unlabeled graph (the implementation is available with the Arabesque release). This is a use case that is favourable to Arabesque as our system is not specifically optimized for match counting.  
The following table compares results of counting three- and four-vertex cliques, using Arabesque and our system (labeled PJ - short for PruneJuice), using the same real-world graphs used for the evaluation of Arabesque in~\cite{Teixeira:2015:ASD:2815400.2815410}. These experiments use the same shared memory machine used earlier. Additionally, for PruneJuice, we present runtimes on 20 compute nodes. (We attempted Arabesque experiments on 20 nodes too, however, Arabesque would crash with the out of memory (OOM) error for the larger \emph{Patent}, \emph{Youtube} and \emph{LiveJournal} graphs. Each compute node in our distributed testbed has 128GB main memory. Our multi-core shared memory testbed, however, has 1.5TB physical memory. Furthermore, for Arabesque, for the workloads that successfully completed on the 20 node deployment, we did not notice any speedup over the single node run.) Note that Arabesque users have to code a purpose-built algorithm for counting cliques, whereas ours and QFrag are generic pattern matching solutions, not optimized to count cliques only. Furthermore, in addition to replicating the data graph, Arabesque also exploits HDFS storage for maintaining the algorithm state (i.e., intermediate matches).

\begin{table}[!t]
{
\footnotesize
\renewcommand{\arraystretch}{1.2}
\captionsetup{font=footnotesize}
\caption{Performance comparison between Arabesque and our pattern matching system (labeled PJ - short for PruneJuice). The table shows the runtime for counting 3-Clique and 4-Clique patterns.  
Here, PruneJuice runtimes for the single node, shared memory are under the column with header PJ (1) while runtimes for the 20 node, distributed deployment are under the column with header PJ (20).}
\label{table:comparison_with_arabesque_cliques}
\begin{center}
\setlength\tabcolsep{2.5pt}
\textcolor{black}{
\begin{tabular}{l|r|r|r|r|r|r}
\hline
\multirow{2}{*}{} & \multicolumn{3}{c|}{3-Clique} & \multicolumn{3}{c}{4-Clique} 
\\
\cline{2-7}
&\multicolumn{1}{c|}{Arabesque} & 
\multicolumn{1}{c|}{PJ (1)} &
\multicolumn{1}{c|}{PJ (20)} &
\multicolumn{1}{c|}{Arabesque} &
\multicolumn{1}{c|}{PJ (1)} &
\multicolumn{1}{c}{PJ (20)} 
\\
\hline
CiteSeer & 3.2s & 0.04s & 0.02s & 3.6s & 0.06s & 0.02s
\\
\hline
Mico & 13.6s & 27s & 11s & 1min & 72min & 21min
\\
\hline
Patent & 1.3min & 17.3s & 1.6s & 2.2min & 32.8s & 8.3s
\\
\hline
Youtube & 6.5min & 2.1min & 12.7s & Crash & 6.4min & 1.4min
\\
\hline
LiveJournal & 8.9min & 2.4min & 11.2s & 2.5hr+ & 1.8hr & 41.3min
\\
\hline
%
\end{tabular}
}
\end{center}
}
\end{table}

PruneJuice was able to count all the clique patterns in all graphs; it took a maximum time of 1.8 hours to count 4-Cliques in the \emph{LiveJournal} graph on the single node, shared memory machine. When using 20 nodes, for the same workload, the runtime came down to 41.3 minutes. Arabesque's performance degrades 
for larger graphs and search templates: Arabesque performs reasonably well for the 3-Clique pattern, for the larger graphs - PruneJuice is at most 3.7$\times$ faster. The 4-Clique pattern, highlights the advantage of our system: for the \emph{Patent} graph, PruneJuice is 4$\times$ faster on the shared memory platform. For the \emph{LiveJournal} graph, Arabesque did not finish in 2.5 hours (we terminate processing). For the \emph{Youtube} graph, Arabasque would crash after runing for 45 minutes. PrinuJuice on the other hand, completed clique counting for both graphs. For the smaller, yet highly skewed \emph{Mico} graph Arabesque outperforms PruneJuice: for the 4-Clique pattern, Arabesque completes clique counting in about one minute, where as it takes PruneJuice 72 minutes on the same platform; this workload highlights the advantage of replicating the data graph for parallel processing which also presents the opportunity for harnessing load balancing techniques that are efficient and effective.


\subsection{Analyzing Sensitivity to Search Template Properties}
\label{sxn:evl_template_sensitivity_analysis}
\begin{figure}[!t]
\centering
\includegraphics[width=4.0in]{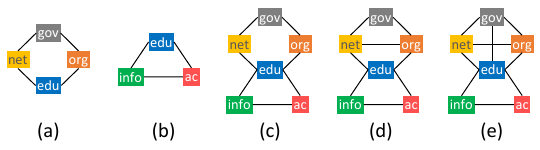} 
\captionsetup{font=footnotesize}
\caption{\textit{WDC} patterns used for template topology sensitivity analysis. Templates (a) and (b) are monocycles, each has a vertex with the label \emph{edu}. Template (c) is created through union of (a) and (b). Templates (d) and (e) are constructed from (c) by incrementally adding one edge at a time.}
\label{fig:wdc_0061}
\end{figure}

\noindent
We investigate the influence of template properties, such as label selectivity and topology, on the runtime of the graph pruning procedure. For this study, we consider the \emph{WDC} graph and the patterns in Fig.~\ref{fig:wdc_0011} and Fig.~\ref{fig:wdc_0061}. 
\\

\textit{\textbf{Impact of Label Selectivity.}}
We consider the WDC-3 and WDC-4 patterns (Fig.~\ref{fig:wdc_0011}): WDC-4, which has the same topology as WDC-3 yet has labels that are less frequent. The two patterns share five out of the eight vertex labels; the labels of WDC-3 and WDC-4, respectively, cover $\sim$15\% and $\sim$4\% of the vertices in the background graph. For WDC-4, the solution subgraph ($|\cV^*|=430$ and $2|\cE^*|=914$) is about two orders of magnitude smaller than that of WDC-3 (see Fig.~\ref{fig:strong_scaling_0011}). The pruning time for WDC-4 is at most 2.6$\times$ faster on 512 nodes, averaging 1.8$\times$ faster across different scales.
\\

\textit{\textbf{Impact of Template Topology.}}
The template topology dictates the type and the number of different constraints to be verified. For example, if the template has a single cycle  (Fig.~\ref{fig:wdc_0061}(a)) then only a single cycle check is required; if the template is not edge-monocyclic (e.g., Fig.~\ref{fig:wdc_0061}(d)) then the relatively more expensive template-driven search is needed for precision guarantees. To understand how the template topology influence performance, we study the \textit{WDC} patterns in Fig.~\ref{fig:wdc_0061}: Templates (a) and (b) each has a single cycle. Template (c) is created through union of (a) and (b). Templates (d) and (e) are constructed from (c) by incrementally adding one edge at a time. Templates (a) -- (c) are edge-monocyclic, thus only need checking cycle constraints. Non-edge-monocyclic templates (d) and (e) require the template-driven search; template (e) needs to verify the existence of a 4-Clique (consisting of vertices with labels \emph{gov}, \emph{org}, \emph{edu}, and \emph{net}). From the topology point of view, among all the constraints here in these examples, the clique is the the most complex substructure and its verification requires the longest walk (TDS constraint). We run these experiments on a 64 node deployment.

\begin{table}[!t]
\footnotesize
\renewcommand{\arraystretch}{1.2}
\captionsetup{font=footnotesize}
\caption{Runtime for pruning (with precision guarantees) and size of the pruned solution subgraph for each \emph{WDC} pattern in Fig.~\ref{fig:wdc_0061}, used for template topology sensitivity analysis. The table lists the number of vertices ($|\cV^*|$) and edges ($2|\cE^*|$) in the solution subgraph.}
\label{table:template_topology_sensitivity}
\centering
\begin{tabular}{c|rrrrr}
\hline
Template & (a) & (b) & (c) & (d) & (e) \\
\hline
{\bf $|\cV^*|$} & 413,527 & 548 & 18,345 & 39 & 8 \\
{\bf $2|\cE^*|$} & 4,095,646 & 1,506 & 139,260 & 166 & 34 \\
Time & 41min & 39s & 2.6min & 2.1min & 1.8min \\
\hline
\end{tabular}
\end{table}

Table~\ref{table:template_topology_sensitivity} lists the runtimes for pruning (with precision guarantees) for the \textit{WDC} patterns in Fig.~\ref{fig:wdc_0061}. The table also shows the number of vertices ($|\cV^*|$) and edges ($2|\cE^*|$) in the solution subgraph for each pattern. While, at first sight, one would expect that the more constraints there are to verify, the slower the system will prune to a precise solution, our experience with the patterns in Fig.~\ref{fig:wdc_0061} proves the contrary.  Template (a) has only one four-cycle to check, however, it has the slowest time-to-solution as it leads to a large solution subgraph (due to the presence of 400M+ vertices in background graph with the labels \emph{org} and \emph{net}). Template (c) and the two templates (d) and (e) that require template-drive search, show, on average, $\sim$20$\times$ faster time-to-solution compared to template (a). The complex templates (c), (d) and (e) introduce additional local and non-local constraints. 
There are at least an order of magnitude more vertices in the background graph, with labels \emph{org} and \emph{net}, that satisfy the constraints of template (a) than that of templates (c), (d) and (e). As a result, templates (c), (d) and (e) eliminate the majority of the non-matching vertices and edges early, leading to a faster time-to-solution; with the most complex template (e) being the rarest and the fastest to finish among the three. 

A key observation here is that it is the abundance of the constraints (in the background graph) that governs performance: template (c), which incorporates the four-cycle constraint that is not present in (b), has an order of magnitude more vertex and edge matches in the background graph, as well as has a slower runtime than that of (b). Similarly, there are only a handful of vertices in the background graph that satisfy the requirements of the complex substructure of (e), i.e., they belong to a clique. Rarity of this constrain leads to rapid pruning, resulting in (e) achieving a faster time-to-solution compared to (c) and (d).


\subsection{Pattern Matching in Graphs with Diverse Topology} 
\label{sxn:evl_graph_sensitivity_analysis}
We demonstrate the ability of effectively processing both labeled and unlabeled graphs with different topological properties: vertex degree distributions, edge density and diameter. To this end, we use there real-world graphs: \emph{Twitter}, \emph{UK Web} and \emph{Road USA} (graph properties are listed in Table~\ref{table:evl_datasets}); and three R-MAT generated graphs (same size yet different vertex degree distribution, Fig.~\ref{fig:graph_degree_distribution_0061}). 
We present runtimes for full match enumeration. We run each experiment on 64 compute nodes; except for the smaller Road USA, for which we use eight compute nodes. 

\subsubsection{Large Real-World Power Law Graphs}
\emph{Twitter} and \emph{UK Web} are billion edge, real-world graphs that have previously used to study a wide range of graph analysis problems; yet, rarely in the context of exact pattern matching. Although both are power law graphs, they have significantly different topologies: Twitter has a more skewed degree distribution, but the larger UK Web graph is denser - it has a higher average vertex degree.

Since \emph{Twitter} and \emph{UK Web} graphs are unlabeled 
we use the same labeling technique used in the past
~\cite{Serafini:2017:QDG:3127479.3131625,Plantenga:2013:ISI:2416443.2416465} -  
we randomly assign vertex labels. For the \emph{Twitter} graph, up to 150 unique labels uniformly distributed among $\sim$41M vertices. For the relatively less skewed \emph{UK Web graph}, we use up to 100 labels. For our experiments, we consider some of the patterns in Fig.~\ref{fig:qfrag_patterns_0011} (previously used by Serafini et al.~\cite{Serafini:2017:QDG:3127479.3131625}). Table~\ref{table:evl_graph_sensitivity_003} lists, for each search template, the full match enumeration time (includes time spent in pruning), match count, and the number of vertices and edges in the solution subgraphs. The results suggest the abundance of acyclic substructures are higher in the \emph{Twitter} graph (10B matches for Q4, compared to 3.8B in the \emph{UK Web} graph). The denser \emph{UK Web} graph has a higher concentration of the cyclic patterns, Q6 and Q8. Q8 is the most abundant - over 45B matches in the background graph; however, the long search duration suggests matches are potentially concentrated within a limited number of graph partitions which limits task parallelism. Similar reasoning applies to long search duration of Q6 in the \emph{Twitter} graph. (We discussed limitations stemming from load imbalance due to such artifacts in \S\ref{sxn:load_balancing_evaluation}.)

\begin{table}[!t]
{
\footnotesize
\renewcommand{\arraystretch}{1.2}
\captionsetup{font=footnotesize}
\caption{\textcolor{black}{Full match enumeration time, for some of the search queries in Fig.~\ref{fig:qfrag_patterns_0011}, in the \emph{Twitter} and \emph{UK Web} graphs (Table~\ref{table:evl_datasets}). 
For each template, the table lists number of vertices ($|\cV^*|$) and edges ($2|\cE^*|$) in the final solution subgraph, match count, and time-to-solution (includes time spent in pruning and match enumeration). 
}}
\label{table:evl_graph_sensitivity_003}
\begin{center}
\setlength\tabcolsep{2.5pt}
\textcolor{black}{
\begin{tabular}{c|rrr|rrr}
\hline
\multicolumn{1}{c|}{} & \multicolumn{3}{c|}{Twitter} & \multicolumn{3}{c}{UK Web} \\
\hline
\#Unique vertex labels & 50 & 100 & 150 & 25 & 50 & 100 \\
\hline
Template & Q4 & Q6 & Q8 & Q4 & Q6 & Q8 \\
\hline
{\bf $|\cV^*|$} & 944K & 91K & 25K & 8M & 1.4M & 230K \\
{\bf $2|\cE^*|$} & 6M & 950K & 338K & 67M & 13M & 2.5M \\
Match count & 10B & 78M & 615M & 3.8B & 2.1B & 45B \\
Time & 1.2hr & 5hr & 1hr & 12.6min & 49.4min & 8.1hr \\
\hline
\end{tabular}
}
\end{center}
}
\end{table}

\begin{table}[!t]
\footnotesize
\renewcommand{\arraystretch}{1.2}
\captionsetup{font=footnotesize}
\caption{\textcolor{black}{Match counting in the \emph{Road USA} graph: the reported runtimes include time spent in pruning as well as match counting. UQ4 has the same topology as Q4 in Fig.~\ref{fig:qfrag_patterns_0011}, however, is unlabeled. The 5-Star pattern is an acyclic graph - a central vertex with four one-hop neighbors (mimicking a four way stop or an intersection). Given the Road USA graph is relatively small, we run these experiments on eight compute nodes.}}
\label{table:evl_graph_sensitivity_004}
\centering
\textcolor{black}{
\begin{tabular}{c|rrrr}
\hline
Unlabeled template & UQ4 & 5-Star & 3-Clique & 4-Clique\\
\hline
Match count & 220M & 66M & 1.3M & 90 \\
Time & 26.7s & 17.3s & 5.0s & 1.6s \\
\hline
\end{tabular}
}
\end{table}

\begin{table}[!t]
{
\footnotesize
\renewcommand{\arraystretch}{1.2}
\captionsetup{font=footnotesize}
\caption{\textcolor{black}{Searches in the three \emph{R-MAT} graphs, in Fig.~\ref{fig:graph_degree_distribution_0061} (a) -- (c), with varying degree distribution, using the Q4 pattern in Fig.~\ref{fig:qfrag_patterns_0011} and the RMAT-2 pattern in Fig.~\ref{fig:rmat_2_0014}. 
For each query in each graph, the table lists the number of vertices ($|\cV^*|$) and edges ($2|\cE^*|$) in the pruned solution subgraph, number of matches and time-to-solution (includes time spent in pruning and match enumeration).}
}
\label{table:evl_graph_sensitivity_005}
\begin{center}
\setlength\tabcolsep{2.5pt}
\textcolor{black}{
\begin{tabular}{c|c|r|r|r}
\hline
\multicolumn{2}{c|}{} & Graph500 &  Chakrabarti et al. & Uniform \\
\hline
\multicolumn{2}{c|}{\#Unique vertex labels} & 26 & 17 & 8 \\
\hline
\multirow{4}{*}{Q4} & {\bf $|\cV^*|$} & 4.4K & 641M & 7.5M\\
& {\bf $2|\cE^*|$} & 7K & 3.5B & 16.4M \\
& Match count & 946 & 4B & 4.2M \\
& Time & 4.2s & 174.2s & 54.1s\\
\hline
%
\multirow{4}{*}{RMAT-2} & {\bf $|\cV^*|$} & 2.3K & 313M & 0 \\
& {\bf $2|\cE^*|$} & 4.1K & 920M & 0 \\
& Match count & 551 & 1.4B & 0 \\
& Time & 8.7s & 83.4s & 52.5s\\
\hline
\end{tabular}
}
\end{center}
}
\end{table}

\subsubsection{Large Diameter Real-World Network}
The \emph{Road USA} graph has a very large diameter and is not labeled. Table~\ref{table:evl_graph_sensitivity_004} lists runtimes for counting matches for four unlabeled patterns. Results show, on the one side, abundance of small acyclic patterns compared to cyclic structures in the road network graph, and on the other side, the ability of our framework to support searches on large, unlabeled graphs with a completely different topology.

\begin{figure}[!h]
\captionsetup[subfigure]{justification=centering}
\centering
\subfloat[Graph 500 - $d_{max}$ 18.7M; \quad\quad  $d_{stdev}$ 1.6K; (0.57, 0.19, 0.19, 0.05)]{\includegraphics[width=2.0in]{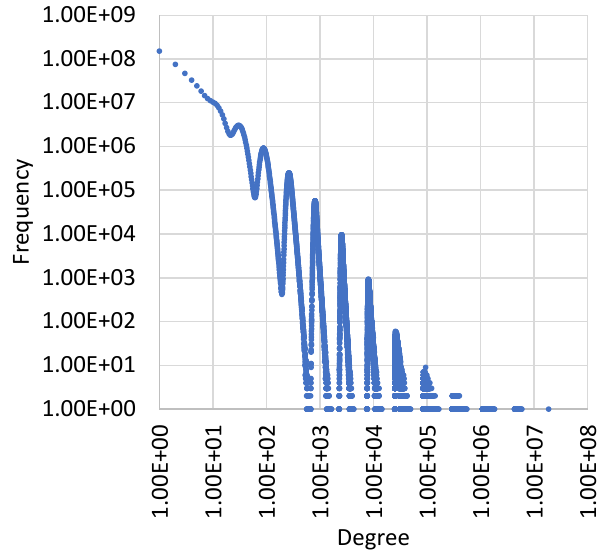}}
\hspace{12pt}
\subfloat[Chakrabarti et al. - $d_{max}$ 48.3K; \quad $d_{stdev}$ 59.3; (0.45, 0.15, 0.15, 0.25)]{\includegraphics[width=2.0in]{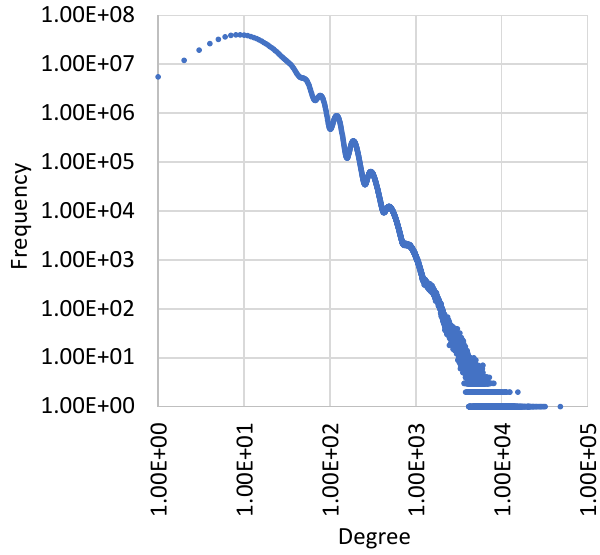}}
\hspace{12pt}
\subfloat[Uniform - $d_{max}$ 482; \quad\quad\quad\quad $d_{stdev}$ 8.6; (0.25, 0.25, 0.25, 0.25)]{\includegraphics[width=2.0in]{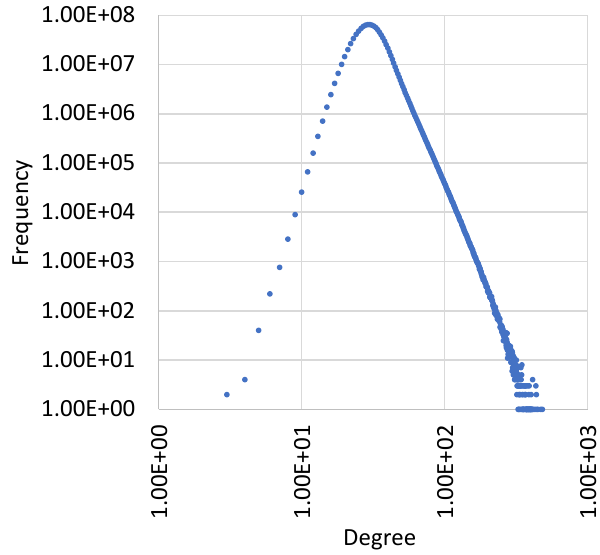}}
\hspace{12pt}
\subfloat[Road USA - $d_{max}$ 9; \quad\quad\quad\quad\quad $d_{avg}$ 2.4; $d_{stdev}$ 0.9]{\includegraphics[width=2.0in]{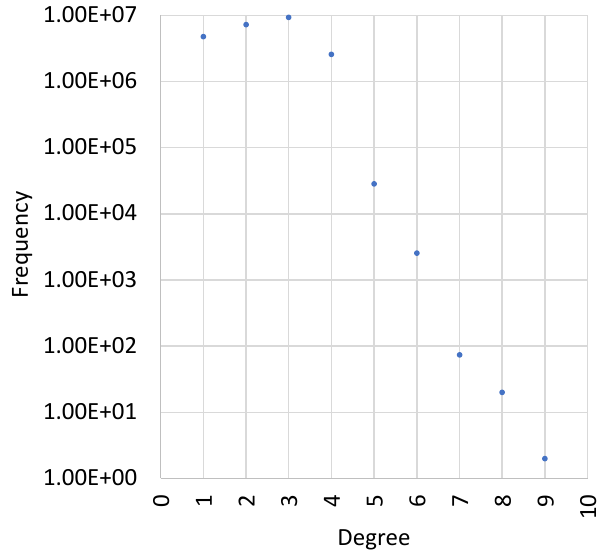}}
\captionsetup{font=footnotesize}
\caption{\textcolor{black}{Figures (a) -- (c) show vertex degree distribution of three R-MAT generated graphs with varying skewness. 
The figures show 
R-MAT edge probabilities $(A, B, C, D)$ used. For (a) -- (c), X and Y axes are in the log scale. Here, the graph Scale (30) and directed edge factor (16) are the same for all three graphs. The storage requirement for each graph is $\sim$270GB. Figure (d) shows vertex degree distribution of the large diameter Road USA graph. Here, only the Y-axis is in log scale.}
\label{fig:graph_degree_distribution_0061}}
\end{figure}

\subsubsection{Experiments on Synthetic Graphs with Varying Vertex Degree Distribution}
We further evaluate our system in presence of graphs with vastly different topologies using R-MAT generated synthetic graphs~\cite{Chakrabarti04r-mat:a}.
To generate power law graphs, the R-MAT model recursively subdivides the graph (initially empty) into four equal-sized partitions and distributes edges within these partitions with predetermined probabilities. Each edge chooses one of the four partitions with probabilities $A$, $B$, $C=B$, and $D$, respectively. These probabilities, determine the skewness of the generated graph: in summary, the higher $A$ is the more skewed the power law distribution becomes. For our experiments, we create three \emph{R-MAT} graphs with the following sets of probabilities: \textit{(i)} The configuration used by the Graph 500 benchmark, (0.57, 0.19, 0.19, 0.05); \textit{(ii)} Parameters suggested by Chakrabarti et al. in~\cite{Chakrabarti04r-mat:a} to simulate real-world scale-free graphs, (0.45, 0.15, 0.15, 0.25); and \textit{(iii)} Equal probability for all four partitions to create graphs with uniform degree distribution (also known as the Erd\H{o}s-R\'enyi graph), (0.25, 0.25, 0.25, 0.25). Fig.~\ref{fig:graph_degree_distribution_0061} (a) -- (c) show the degree distribution of these \emph{R-MAT} graphs. For all the graphs, we use the same Scale (30) and (directed) edge factor (16) - leading to (undirected) graphs with 34B edges. We follow the same approach as used for weak scaling experiments in~\cite{Reza:2018:SC:PM}, \S5A, to generate vertex labels: we use vertex degree information to create vertex labels, computed using the formula, $\ell(v_i) = \lceil \log_2(d(v_i)+1)\rceil$.

Table~\ref{table:evl_graph_sensitivity_005} shows results for full match enumeration in the three \emph{R-MAT} graphs for the following two queries: Q4 in Fig.~\ref{fig:qfrag_patterns_0011} and RMAT-2 in Fig.~\ref{fig:rmat_2_0014}. 
For both search templates, most matches are found for the configuration labeled Chakrabarti et al. This is because the graph in Fig.~\ref{fig:graph_degree_distribution_0061}(b) has a higher frequency of high-degree vertices compared to the graphs in Fig.~\ref{fig:graph_degree_distribution_0061}(a) and (c); since we use degree based vertex labels, this has direct impact on match density. Although, for the smaller Q4 pattern, the \emph{R-MAT} graph with uniform degree distribution has more matches than the Graph 500 configuration; in the uniform graph, no matches were found for the larger seven vertex RMAT-2 pattern with unique labels: low variance in degree distribution means 99.9\% of the graph vertices have one of the top four most frequent labels.


\section{Limitations}
\label{sxn:limitations}

\noindent
We categorize the limitations of our proposed system based on their respective sources. 
\\

\textit{\textbf{Limitations stemming from major design decisions.}} 
Our pipeline inherits the limitations of systems primarily designed for exact matching (compared to systems that trade accuracy for performance, e.g., based on sampling~\cite{Iyer:2018:ASAP:222637} or graph simulation~\cite{Fan:2010:GPM:1920841.1920878}). Similarly, our system inherits all limitations of its communication and middleware infrastructure, MPI and HavoqGT, respectively. One example is the lack of sophisticated flow/congestion control mechanism provided by these infrastructures which sometimes lead to message buildup and system collapse.
\\

\textit{\textbf{Limitations stemming from the targeted uses cases.}} 
In the same vein, we note that our system targets a graph analytics scenario (queries that need to cover the entire graph), rather than the traditional graph database queries that attempt to find a specific pattern around a vertex indicated by the user.  
\\

\textit{\textbf{Limitations stemming from attempting to design a generic system.}} 
Systems optimized for specific patterns may perform better, e.g., systems optimized to count/enumerate triangles~\cite{Suri:2011:CTC:1963405.1963491}, treelets~\cite{Zhao.2012.SAHAD.6267876} 
or systems relying on multi-join indices~\cite{Sun:2012:ESM:2311906.2311907} to support patterns with limited diameter.


\section{Conclusion}
\label{sxn:conclusion}

\noindent
This paper presents a new algorithmic pipeline to support pattern matching on large-scale metadata graphs on distributed memory systems. We capitalize on the idea of \textit{graph pruning} via \textit{constraint checking} and develop asynchronous algorithms that use both vertex and edge elimination to iteratively prune the original graph and reduce it to a subgraph which is the union of all matches. We have developed pruning techniques that \textit{guarantee a solution with} 100\% \textit{precision} (i.e., no false positives in the pruned subgraph) and 100\% \textit{recall} (i.e., all vertices and edges participating in matches are included) for \textit{arbitrary search patterns}. Our algorithms are \textit{vertex-centric} and \textit{asynchronous}, thus, they map well onto existing high-performance graph frameworks. Our evaluation using up to 257 billion edge real-world graphs and up to 4.4 trillion edge synthetic \textit{R-MAT} graphs, on up to 1,024 nodes (36,864 cores), confirms the scalability of our solution. We demonstrate that, depending on the search template, our approach prunes the graph by orders of magnitude which enables match enumeration and counting on graphs with trillions of edges. Our success stems from a number of key design ingredients: asynchronicity, aggressive vertex and edge elimination while harnessing massive parallelism, intelligent work aggregation to ensure low message overhead, effective pruning constraints, and lightweight per-vertex state.

While we believe our system, as described, represents a significant advance in practical pattern matching in large, real-world graphs, further investigations in a number of areas can improve the efficiency and robustness of our solution. \textit{(i)} The graph pruning pipeline introduces a number of decision problems. At present it uses ad-hoc heuristics, developed based on our intuition. We believe modelling approaches similar to the one explored in~\cite{Nicolas:2018:IA3:PM:8638389}, can be used to inform the following decisions: constraint selection and ordering, when to trigger load balancing and when to switch from pruning to direct enumeration. \textit{(ii)} The current prototype implementation can be extended to enable support for a richer set of subgraph matching scenarios, e.g., pattern matching in graphs and templates with edge metadata; querying dynamic/time-evolving graphs~\cite{Boldi:2008:LTW:1480506.1480511,Sallinen:2016:GCC:3014904.3014945, Han:2014:CGE:2592798.2592799, Vora:2017:KFA:3093337.3037748} and approximate pattern matching~\cite{Bunke:1983:IGM:2305869.2306079,Alon:2008:BNM:1388083.1388101,CONTE.2004.TYGMPR.doi:10.1142.S0218001404003228,Iyer:2018:ASAP:222637}. \textit{(iii)} Further design/system optimizations, especially for non-local constraint checking and full match enumeration, to improve memory and message efficiency, load balance and task parallelism.



\bibliographystyle{ACM-Reference-Format}
\bibliography{references_6}


\end{document}